\documentclass[12pt,a4paper]{article}
\pdfoutput=1

\usepackage{jcappub}
\allowdisplaybreaks

\usepackage[margin=1in]{geometry}
\usepackage{graphicx}
\usepackage{amsfonts}
\usepackage{subfigure}
\usepackage{float}
\usepackage{hyperref}
\usepackage[dvipsnames]{xcolor}        
\usepackage{amsmath}
\usepackage[normalem]{ulem} 
\usepackage{empheq}
\usepackage{amssymb}
\usepackage{tikz}
\usepackage{tabularx}
\usepackage{booktabs}

\definecolor{redViolet}{rgb}{0.8,0.0,0.4}

\def\lsim{~\rlap{$<$}{\lower 1.0ex\hbox{$\sim$}}}
\def\bsim{~\rlap{$>$}{\lower 1.0ex\hbox{$\sim$}}}

\def\ln{{\rm ln}}

\def\ka{\kappa}

\def\be{\begin{equation}}
\def\ee{\end{equation}}
\def\bea{\begin{eqnarray}}
\def\eea{\end{eqnarray}}
\def\ba{\begin{align}}
\def\ea{\end{align}}
\def\bi{\begin{itemize}}
\def\ei{\end{itemize}}

\newcommand{\deta}{r}
\newcommand{\Hcal}{\mathcal H}

\newcommand{\Gaunt}[6]{{\cal G}^{#1#2#3}_{#4#5#6}}

\newcommand{\de}{\partial}

\newcommand{\bn}{{\mathbf n}}

\def\d{{\rm d}}

\newcommand{\nn}{ \nonumber }

\newcommand{\fnl}{f_{\rm NL}}
\newcommand{\fnleq}{f_{\rm NL}^{\rm eq.}}
\newcommand{\fnlloc}{f_{\rm NL}^{\rm loc.}}
\newcommand{\fnlorth}{f_{\rm NL}^{\rm orth.}}

\def\mathbi#1{\textbf{\em #1}}

\def\vk{\mathrm{\bf k}}

\def\vx{\mathrm{\bf x}}

 \def\vx{\vec{ x}}
   
\def\d{\rm d}

\def\d{{\rm d}}
\def\Om{\Omega}

\def\De{\Delta}

\def\vx{\bf{x}}

\def\bx{{\bf{x}}}
\def\bk{{\bf{k}}}

\def\bv{{\bf{v}}}

\def\ba{{\vec{g}}}

\newcommand{\HH}{\mathcal{H} }

\def\grad{\mathbi{$\nabla$}}

\title{Non-Gaussianities due to Relativistic Corrections to the Observed Galaxy Bispectrum}

\author[a,b]{E.~Di~Dio,}
\author[c]{H.~Perrier,}
\author[c]{R.~Durrer,}
\author[d]{G.~Marozzi,}
\author[c,e]{A.~Moradinezhad~Dizgah,}
\author[f]{J.~Nore\~{n}a,}
\author[c]{A.~Riotto}
\affiliation[a]{INAF - Osservatorio Astronomico di Trieste, Via G. B. Tiepolo 11, I-34143 Trieste, Italy}
\affiliation[b]{INFN, Sezione di Trieste,
Via Valerio 2, I-34127 Trieste, Italy}
\affiliation[c]{University of Geneva, Department of Theoretical Physics and Center for Astroparticle Physics (CAP),
24 quai E. Ansermet, CH-1211 Geneva 4, Switzerland}
\affiliation[d]{Centro Brasileiro de Pesquisas F\'{i}sicas, Rua Dr. Xavier Sigaud 150, Urca, CEP 
22290-180, Rio de Janeiro, Brazil}
\affiliation[e]{Department of Physics, Harvard University, 17 Oxford St., Cambridge, MA 02138, USA}
\affiliation[f]{Instituto de F\'{i}sica, Pontificia Universidad Cat\'{o}lica de Valpara\'{i}so, Casilla 4059, Valpara\'{i}so, Chile}
\emailAdd{Enea.DiDio@oats.inaf.it}
\emailAdd{Hideki.Perrier@unige.ch}
\emailAdd{Ruth.Durrer@unige.ch}
\emailAdd{Marozzi@cbpf.br}
\emailAdd{amoradinejad@physics.harvard.edu}
\emailAdd{jorge.norena@pucv.cl}
\emailAdd{Antonio.Riotto@unige.ch}

\abstract{
High-precision constraints on primordial non-Gaussianity (PNG) will significantly improve our understanding of the physics of the early universe. Among all the subtleties in using large scale structure observables to constrain PNG, accounting for relativistic corrections to the clustering statistics is particularly important for the upcoming galaxy surveys covering progressively larger fraction of the sky. We focus on relativistic projection effects due to the fact that we observe the galaxies through the light that reaches the telescope on perturbed geodesics. These projection effects can give rise to an effective $f_{\rm NL}$ that can be misinterpreted as the primordial non-Gaussianity signal and hence is a systematic to be carefully computed and accounted for in modelling of the bispectrum. We develop the technique to properly account for relativistic effects in terms of purely observable quantities, namely angles and redshifts. We give some examples by applying this approach to a subset of the contributions to the tree-level bispectrum of the observed galaxy number counts calculated within perturbation theory and estimate the corresponding non-Gaussianity parameter, $f_{\rm NL}$, for the local, equilateral and orthogonal shapes. For the local shape, we also compute the local non-Gaussianity resulting from terms obtained using the consistency relation for observed number counts. Our goal here is not to give a precise estimate of $f_{\rm NL}$ for each shape but rather we aim to provide a scheme to compute the non-Gaussian contamination due to relativistic projection effects. For the terms considered in this work, we obtain contamination of $f_{\rm NL}^{\rm loc} \sim {\mathcal O}(1)$.}

\begin{document}
\maketitle

\section{Introduction}

Understanding the mechanism behind the generation of primordial fluctuations which sets the seed of the Large Scale Structure (LSS) of the Universe, is one of the main open fundamental questions in cosmology. Latest high-precision measurements of temperature and polarization anisotropies of the Cosmic Microwave Background (CMB) from Planck satellite \cite{Ade:2015ava} are consistent with adiabatic, nearly scale-invariant and Gaussian primordial fluctuations in agreement with the predictions of simplest inflationary models.  Constraints on the level of non-Gaussianity of the primordial fluctuations will result in significant improvement in our understanding of the physics of the early universe. For example, achieving a high sensitivity on the local non-Gaussianity parameter $\fnlloc$ is of great theoretical interest since it could enable us to distinguish between various models of inflation \cite{Bartolo:2004if}.

High-precision measurements of the clustering statistics of LSS such as power spectrum and bispectrum are our best hope to obtain such an accuracy, beyond the limits set by the CMB, due to the large number of available modes. Indeed, the target accuracy for the upcoming LSS surveys is  $\Delta \fnlloc \simeq {\mathcal O}(1)$ for local primordial non-Gaussianity (PNG) \cite{Dore:2014cca,Tellarini:2016sgp}, while for other shapes is $\Delta \fnl \sim \mathcal{O}(10)$ \cite{Baldauf:2016sjb}. PNG of the local shape leaves a distinct signature on the galaxy power spectrum by inducing a scale-dependant contribution to the biasing relation between the dark matter and the galaxy overdensities \cite{Dalal:2007cu,Matarrese:2008nc}. This signature has been used to set constraint on $f_{\rm NL}$ of the local shape \cite{Afshordi:2008ru,Slosar:2008hx}. The galaxy bispectrum contains additional information on PNG since it is sensitive to all shapes. It also provides a consistency check for the constraints obtained with the power spectrum \cite{Scoccimarro:2003wn,Sefusatti:2007ih,Jeong:2009vd}.
On the other hand, compared to the CMB, constraining $f_{\rm NL}$ from LSS and fully exploiting its potential is more challenging.

 Accurate theoretical modelling of these observables should necessarily account for four factors. The first is the non-linearities in the evolution of the dark matter density field. Unlike the CMB where the anisotropies are small enough to be studied in the linear regime of cosmological perturbations, the LSS is highly non-linear due to nonlinear growth of structures. Second, the LSS is a biased tracer of the underlying dark matter. This relation in principle is non-linear and scale-dependant. Third, the LSS biased tracers are observed in redshift space rather than in real space: their peculiar velocity unavoidably induces redshift-space distortions (RSD) in their observed overdensity. Fourth, which is the focus of the current paper, on the largest scales probed by upcoming LSS surveys, one has to take into account  relativistic projection effects which affect the observed redshift and angular position of the galaxy, beyond RSD. Furthermore, even at smaller scales when correlating different redshift bins lensing effects can give an important contribution.

Analogously to what is done for the CMB, any contribution to the bispectrum which looks like PNG should be subtracted from the measurement before estimating the level of the primordial signal. In the case of the CMB, the main contamination to $\fnlloc$ is  found to be the ISW-lensing effect \cite{Lewis:2012tc,Kim:2013nea}. For the galaxy bispectrum however the task in hand is more complicated as even at the tree-level in perturbation theory, 
the signal is dominated by non-primordial contributions. 
Therefore, to achieve the target sensitivity of $\Delta f_{\rm NL} \simeq {\mathcal O}(1)$ for the upcoming LSS surveys, determining the contamination to the non-Gaussian parameter at the same accuracy is essential.

In addition to the well-known contributions from the gravitational evolution often referred to as ``Newtonian terms'', which can be calculated using a Newtonian perturbation theory scheme such as standard perturbation theory (SPT), there are several relativistic projection contributions. These corrections arise due to the fact that we observe galaxies through photons which have travelled on perturbed geodesics in a clumpy universe into our telescope \cite{Yoo:2010ni,Bonvin:2011bg,Challinor:2011bk,Jeong:2011as}. In computing the observed quantities such as galaxy number counts, we therefore need to account for this projection\footnote{Note that the contribution of the projection effects should not be confused with the bispectrum generated by the dynamics beyond the Newtonian approximation as in \cite{Bartolo:2005xa,Bartolo:2010rw,Villa:2014foa}.}on our past 
light-cone. 

Also it should be noted that the Newtonian terms are known to give a large contribution to local non-Gaussianity of the order of $10^3-10^4$ \cite{Verde:1999ij,Bartolo:2005xa}. In this sense, the effects we consider here are sub-leading corrections to the Newtonian terms. However, if ignored, they can bias the measurement of the primordial signal by a value comparable to the precision promised by future surveys. Therefore, they need to be properly taken into account in the modelling be it analytically or using relativistic N-body simulations, see e.g.~\cite{Adamek:2015eda}.

Within perturbation theory, in order to calculate the tree-level bispectrum induced by projection effects and assuming Gaussian initial conditions, a fully relativistic description of the observed galaxy number counts up to second order is necessary.  This has been recently developed by several groups \cite{Bertacca:2014dra,Yoo:2014sfa,DiDio:2014lka} for the observed galaxy number counts in terms of the observed angles and redshifts. The bispectrum evaluated within the perturbative scheme has the advantage that it is complete within the validity region of perturbation theory.
Also it is valid for any triangular configuration. However estimating the effective contamination to  $f_{\rm NL}$ using the full expression is rather cumbersome as there are hundreds of terms contributing. Up to now there has been no estimate of the effective $f_{\rm NL}$ from the second-order calculation.

Taking a different approach, Kehagias et al.~\cite{Kehagias:2015tda} derived a consistency relation for the observed galaxy bispectrum which allows to calculate the observed galaxy bispectrum in the squeezed limit neglecting contributions from  curvature and tidal effects which are proportional to second derivatives of the gravitational potential. Since the bispectrum of the local-shape is assumed to peak at the squeezed limit, the authors used this result to give an order of magnitude estimation of the effective $f_{\rm NL}$ due to relativistic projection effects that has to be subtracted from the measured value. 
On the one hand, this relation is more tractable than the full second-order calculation and can be used at arbitrary small scales for the short modes. On the other,  it is only valid in the squeezed limit by construction and there are physical effects such as curvature and tidal contributions that can not be captured by this argument. 

Another approach has recently been taken in Umeh et al.~\cite{Umeh:2016nuh}. There they compute the bispectrum including all  terms of the relativistic calculation in~\cite{Bertacca:2014dra} which do not contain integrals over the line of sight. But they present the result in Fourier space and not in the observable angular and redshift space. Also their definition of $f_{\rm NL}$ differs from ours which makes it difficult to directly compare the results.

The fully relativistic  approach and the calculation using the  consistency relation should agree on regimes where both assumptions, namely validity of perturbation theory and super-Hubble long modes, are valid. We leave this interesting comparison to a future work, while we are now interested in developing and applying  a proper scheme to include relativistic effects in terms of directly observable quantities.

In particular, we evaluate here the level of non-Gaussianity generated by a subset of the contributions to the observed galaxy bispectrum calculated in~\cite{DiDio:2014lka,DiDio:2015bua}. We classify various corrections by the number of derivatives with respect to the metric perturbations and integrals along the light-cone they contain.
Here, what we mean by the non-Gaussianity induced by projection effects is the systematic shift in a measurement of $f_{\rm NL}$ if these effects are neglected.
The goal of the paper is not to make an exhaustive analysis of all (hundreds of) terms appearing in the second order expression to give a precise value of the effective $f_{\rm NL}$ due to relativistic corrections. Because cancellations between terms can happen, a full analysis will very likely be required in the end and we leave it for future work. Here we rather aim at highlighting the importance of  projection effects by accurately estimating the level of non-Gaussianity that some of them can generate. It should be emphasized that this calculation is the first of its kind; we develop the numerical scheme to evaluate the bispectrum templates for the local, equilateral and orthogonal shapes in redshift and angular space: $z \ell m$-space. This is a necessary ingredient in accurately estimating the value of $f_{\rm NL}$ of a given non-Gaussian shape, as one should project the theoretical bispectrum calculated in terms of the observed angles and redshifts onto the non-Gaussian 
templates in $z \ell m$-space and sum over all the triangular configurations. 

In this preliminary work, we provide simple examples essentially limited to a single redshift slice and to relatively large scales. By including several redshift bins it is possible to recover the full 3-dimensional information, as shown for the redshift dependent angular power spectra in~\cite{Asorey:2012rd,DiDio:2013sea}. Therefore redshift evolution and smaller scales may help in disentangling primordial non-Gaussianity from the LSS contaminations.

The rest of this paper is organized as follows: in Sect.~\ref{secng} we set up the basics by reviewing bispectrum templates in Fourier space. Next we calculate their counterpart in $z \ell m$-space and point out some subtleties in  the numerical evaluation of the integrals. We close this section by defining the shape correlation in $z \ell m$-space between local and equilateral templates and we define the $f_{\rm NL}^{\rm eff}$ as the normalized amplitude of the projected bispectrum of the projection effects onto a particular PNG template. In Sect.~\ref{secgrcorr}, we review the theoretical bispectrum from relativistic corrections as calculated via consistency relation argument and also the subset of terms in the bispectrum calculated from second-order perturbation theory. In Sect.~\ref{secresults} we present the estimates for the effective $f_{\rm NL}$ from various terms that we considered for local, equilateral and orthogonal shapes. We also estimate $f_{\rm NL}$ for the local shape using  the bispectrum obtained from consistency relation. Finally, we conclude in Sect.~\ref{secconclusion}.

All numerical results presented have been obtained with the following cosmological parameters: $h= 0.67$, $\omega_b = 0.022 $, $\omega_{cdm} =0.12$ and vanishing curvature. The primordial curvature power spectrum has the amplitude $A_s = 2.215 \times 10^{-9} $, the pivot scale $k_\text{pivot} = 0.05 \text{Mpc}^{-1}$, the spectral index $n_s = 1$ and no running.

%%%%%%%%%%%%%%%%%%%%%%%%%%%%%%%%%%%%%%%%%%%%%%%%%

\section{Primordial non-Gaussianities in the observed bispectrum}\label{secng}
To quantify the contamination from  projection effects to the measured primordial $f_{\rm NL}$ of a given shape, we need to project the bispectrum from relativistic corrections onto the corresponding primordial bispectrum template for that shape. For the rest of our discussion, we focus on local, equilateral and orthogonal templates. Below we first set the basic notation and define the initial conditions in terms of the primordial bispectrum for these PNG shapes. Next we derive corresponding templates in terms of galaxy number counts in $z\ell m$-space and discuss in details some subtleties in evaluating the numerical integrals. 

\subsection{Primordial non-Gaussian shapes}
While the statistics of Gaussian fluctuations is fully determined by the lowest order correlation function, i.e.~2-point function or its Fourier transform, the power spectrum, to describe non-Gaussian fields one needs to consider higher order correlation functions. For the rest of our discussion we focus on the lowest order non-Gaussian statistics of primordial fluctuations, the 3-point function and its counterpart in Fourier space, the bispectrum. We therefore consider the non-Gaussian initial conditions given in terms of the bispectrum of the primordial curvature perturbation $\zeta$ defined as
\be
\langle \zeta(\bk_1)  \zeta(\bk_2) \zeta(\bk_3)\rangle = (2\pi)^3 \delta_D^{(3)}(\bk_{1}+\bk_{2}+\bk_{3}) B_\zeta(k_1,k_2,k_3) \, ,
\ee
where $\langle ... \rangle$ denotes the ensemble average and we assumed statistical homogeneity and isotropy.
The primordial bispectrum is often parameterized as
\be  
B_\zeta(k_1,k_2,k_3) = f_{\rm NL}^{\rm shape}  F_{\rm shape}(k_1, k_2, k_3) \, ,
\ee
where $f_{\rm NL}^{\rm shape}$ is a normalized amplitude and $F_{\rm shape}(k_1,k_2,k_3)$ encodes the functional dependance of the primordial bispectrum on the specific triangular configuration. Therefore constraints on PNG are reported as constraints on the amplitude parameter for a particular shape. Different inflationary models, give specific prediction for the amplitude and the shape of bispectrum \cite{Chen:2010xka,Liguori:2010hx}. Therefore constraints on PNG provide an invaluable window to distinguish between inflationary models. 

For the rest of our discussion, we consider three commonly used separable shapes\footnote{As we will see in the next sections, being separable in Fourier space does not imply to be separable in terms of the observable bispectrum in redshift and angular space.}, the local, equilateral and orthogonal. For a scale-invariant primordial power spectrum $P_\zeta(k)= A_\zeta k^{-3}$, the local template is given by \cite{Gangui:1993tt,Wang:1999vf, Verde:1999ij, Komatsu:2001rj}
\be\label{Floc}
B_\zeta^\text{local}(k_1, k_2, k_3)  = \frac{6}{5} \fnlloc  \left[P_\zeta(k_1)P_\zeta(k_2) + 2 \ {\rm perms.} \right] 
 = \frac{6}{5}\fnlloc A_\zeta^2 \left( \frac{1}{k_2^3k_3^3}   + 2 \ {\rm perms.}   \right) \,,
\ee
while the equilateral template is defined as \cite{Babich:2004gb}
\be\label{Fequil}
B_\zeta^\text{equil}(k_1, k_2, k_3)  = \frac{18}{5}  \fnleq A_\zeta^2 \left[-\left( \frac{1}{k_1^3 k_2^3} + 2 \ {\rm perms.} \right) -\frac{2}{k_1^2 k_2^2 k_3^2}  + \left( \frac{1}{k_1 k_2^2 k_3^3} + 5 \,{\rm perms.} \right) \right] \, , 
\ee
and the orthogonal template is given by \cite{Senatore:2009gt}
\be\label{Fortho}
B_\zeta^\text{orth}(k_1, k_2, k_3)  = \frac{18}{5} \fnl^{\rm orth} A_\zeta^2 \left[-\left( \frac{3}{k_1^3 k_2^3} + 2 \ {\rm perms.} \right) -\frac{8}{k_1^2 k_2^2 k_3^2}  + \left( \frac{3}{k_1 k_2^2 k_3^3} + 5 \,{\rm perms.} \right) \right] \, .
\ee

%%%%%%%%%%%%%%%%%%%%%%%%%

\subsection{Computing the non-Gaussian shapes in $z\ell m$-space} \label{sec:num_int}
To obtain the shape templates in terms of observable quantities we need to compute the bispectrum related to the 3-point function 
\be
\langle \Delta_g \left( \bn_1 ,z_1 \right)  \Delta_g \left(  \bn_2 ,z_2 \right)  \Delta_g \left(  \bn_3 ,z_3 \right) \rangle
\,,
\ee
given the PNG initial conditions set by Eqs.~(\ref{Floc}-\ref{Fortho}), and the observable galaxy number counts, in terms of the observed redshift $z$ and the angular position\footnote{Following the notation of~\cite{Bonvin:2011bg,DiDio:2013bqa}, the unit vector $\bn$ denotes the direction of light propagation, from the source to the observer.} $\bn$, defined as
\be
\Delta_g \left( \bn , z\right) = \frac{n_g \left( \bn , z \right) -\langle n_g \rangle \left( z \right)}{\langle n_g \rangle \left( z \right)}\,,
\ee
where $\langle ... \rangle$ denotes the angular mean at fixed observed redshift $z$ and $n_g \left( \bn ,z \right) $ is the number density of sources per redshift and per solid angle. For non-Gaussian initial conditions, the lowest order non-vanishing contribution to the 3-point function is given by the linear galaxy number counts~\cite{Challinor:2011bk,Bonvin:2011bg,DiDio:2013bqa}
\bea \label{number_counts}
 \Delta_g^{(1)} \left( \bn, z \right) &=& b \delta^\text{sync} +(5s -2)\Phi + \Psi + \frac{1}{\HH}
\left[\dot \Phi +\partial^2_r v \right] +  \left( f_\text{evo} - 3 \right) \HH v\nonumber \\  &&
+ \left(\frac{{\dot\HH}}{\HH^2}+\frac{2-5s}{r_S\HH} +5s-f_{\rm evo}\right)\left(\Psi+\partial_r v+ 
 \int_0^{r_S}\hspace{-0.3mm}dr(\dot \Phi+\dot \Psi)\right) 
   \nonumber \\  &&  \label{DezNF}
+\frac{2-5s}{2r_S}\int_0^{r_S}\hspace{-0.3mm}dr \left[2-\frac{r_S-r}{r}\Delta_\Om\right] (\Phi+\Psi) \, ,
\eea
where $\Psi$ and $\Phi$ denote the Bardeen potentials\footnote{In our convention the Bardeen potentials $\Psi$ and $\Phi$ correspond, respectively, to the time and spatial component of the metric perturbations in Newtonian gauge (see Appendix~\ref{explicit_fullconsist}). We warn the reader that in \cite{Kehagias:2015tda} the opposite sign convention has been used.}, $v$ the (gauge invariant) velocity potential in Newtonian gauge, related to the peculiar dark matter velocity through $\bv = -\grad v$, and $\HH=\dot a/a$ is the comoving Hubble parameter. We denote with a dot the partial derivative with respect to conformal time $\eta$ and $\Delta_\Omega$ denotes the angular part of the Laplacian. The dark matter perturbations are related to galaxies number counts through some bias parameters: $b$ is the linear galaxy bias which acts on the matter perturbation in synchronous gauge\footnote{We apply the galaxy bias $b$ in synchronous comoving gauge. This choice is well justified from the assumption that galaxies and the underlying dark matter density field experience the same gravitational field and therefore they move with the same velocity~\cite{Baldauf:2011bh,Jeong:2011as}. Hence in their rest frame we can apply the linear bias prescription.}, while $s$ and $f_\text{evo}$ are, respectively, the magnification bias and the evolution bias defined as
 \be
 s\equiv \frac{{\partial \ln} \ \bar n_g}{{\partial \ln} \ L},    \qquad \qquad f_{\rm evo} \equiv \frac{{\partial \ln} (a^3 \bar n_g)}{\mathcal H \partial\eta},
\ee
where we denote with an overbar the background quantities and $L$ is the threshold luminosity of a given galaxy survey.
Since in this work we limit our analysis to a single redshift bin, for sake of simplicity we will not consider the terms integrated along the line of sight of Eq.~\eqref{number_counts}. We remind that we adopt this approximation only in computing the PNG templates in $z\ell m$-space, and not for the physical relativistic bispectrum. 
In the rest of the paper we consider the following bias parameters $b=1$, $s=0$ and $f_\text{evo}=0$.

To compute the redshift dependent angular bispectra, we expand the galaxy number counts in spherical harmonics $Y_{\ell m}$ 
\be \label{spherical_expansion}
\Delta_g\left( \bn , z \right) = \sum_{\ell m} a_{\ell m }^{\Delta_g} \left( z \right) Y_{\ell m } \left( \bn \right)\,,
\ee
whose coefficients are given by
\be
a_{\ell m}^{\Delta_g}(z)=\int d \Omega_\bn \ Y_{\ell m}^*({\bn}) \Delta_g \left( \bn ,z \right) \, .
\ee
Following~\cite{Bonvin:2011bg}, it is straightforward to rewrite the spherical harmonics coefficients, for the terms of Eq.~\eqref{number_counts} considered, as 
\be
a_{\ell m}^{\Delta_g}(z)= i^\ell\frac{4 \pi}{(2\pi)^3}\int d^3 k \ Y_{\ell m}^*(\hat{\bk}) \Delta_\ell(k,z)\zeta(\bk)
\ee
with
\be \label{numbercounts_transfer}
\Delta_\ell(k,z) = \sum_i \Delta_\ell^i(k,z)\,,
\ee
where the sum is over different angular transfer functions defined in Appendix A.4 of~\cite{DiDio:2013bqa}.

The bispectrum is then given by
\bea
&&
 B_{\ell_1\ell_2\ell_3}^{m_1 m_2 m_3}(z_1,z_2,z_3)  =  \left< a_{\ell_1 m_1}^{\Delta_g}(z_1) a^{\Delta_g}_{\ell_2 m_2}(z_2) a^{\Delta_g}_{\ell_3 m_3}(z_3) \right>   \nonumber \\
& =& i^{\ell_1+\ell_2+\ell_3}  \int \prod_{p=1}^3 \left \lbrace d^3 k_p \frac{4 \pi}{(2\pi)^3} Y^*_{\ell_p m_p}(\hat{\bk}_p) \Delta_{\ell_p}(k_p,z_p) \right \rbrace \langle \zeta(\bk_1) \zeta (\bk_2)\zeta(\bk_3)\rangle \nonumber   \\
& =& i^{\ell_1+\ell_2+\ell_3}  \int \prod_{p=1}^3 \left\lbrace d^3 k_p \frac{4 \pi}{(2\pi)^3} Y^*_{\ell_p m_p}(\hat{\bk}_p) \Delta_{\ell_p}(k_p,z_p) \right\rbrace    (2\pi)^3 \delta_D^{(3)}(\bk_{1}+\bk_{2}+\bk_{3}) 
\nonumber \\
& & \times
\fnl^{\rm shape} F_{\rm shape} (k_1, k_2, k_3) \,.
\eea
One then expands the delta-Dirac function
\bea
&& \delta_D^{(3)}(\bk_1+\bk_2+\bk_3) =  \int d^3 x \, \frac{1}{(2\pi)^3} \,  e^{i (\bk_1 + \bk_2 + \bk_3) \cdot {\vx}} \nonumber \\
&=& \frac{1}{(2\pi)^3} \int \, dr r^2 d \Omega_{{\bn}}\, e^{- i (\bk_1 + \bk_2 + \bk_3) \cdot r {\bn}} \nonumber \\
&=& \frac{1}{(2\pi)^3} \int \, dr r^2 d \Omega_{{\bn}}\, \prod_{q=1}^3 \left\lbrace 4 \pi \sum_{\ell'_q=0}^\infty \sum_{m'_q=-\ell'_q}^{\ell'_q} (-i)^{\ell'_q} j_{\ell'_q}(r k_q)Y_{\ell'_q m'_q}(\hat{\bk}_q) Y^*_{\ell'_q m'_q}({\bn}) \right\rbrace \label{delta} \,,
\eea
and, by performing the angular integrals over $d\Omega_{\hat\bk_i}$ and $d\Omega_{\bn}$, we obtain
\bea \label{full}
B_{\ell_1 \ell_2 \ell_3}^{m_1 m_2  m_3} &=&
\Gaunt{m_1}{m_2}{m_3}{\ell_1}{\ell_2}{\ell_3} \left(\frac{2}{\pi}\right)^3
 \int  \d k_1  \d k_2  \d k_3  \int dr r^2   \prod_{p=1}^3 \left\lbrace  k_p^2   \Delta_{\ell_p}(k_p,z_p) 
 j_{\ell_p}(r k_p) \right\rbrace 
 \nonumber \\
 &&
\qquad \qquad \qquad \qquad \qquad \times
  \fnl^{\rm shape} F_{\rm shape}(k_1, k_2, k_3) \,,
   \eea
where we have introduced the Gaunt factor 
\bea
 \Gaunt{m_1}{m_2}{m_3}{\ell_1}{\ell_2}{\ell_3}  &=& \int d\Omega_\bn Y_{\ell_1 m_1 } \left( \bn \right)  Y_{\ell_2 m_2 } \left( \bn \right)  Y_{\ell_3 m_3 } \left( \bn \right) 
\nonumber \\
 &=& \left( \begin{array}{ccc} \ell_1 & \ell_2 & \ell_3 \\ 0 & 0 & 0 \end{array} \right)  \left( \begin{array}{ccc} \ell_1 & \ell_2 & \ell_3 \\ m_1 & m_2 & m_3 \end{array} \right) \sqrt{\frac{\left( 2 \ell_1+1 \right) \left( 2 \ell_2+1 \right) \left( 2 \ell_3+1 \right)}{ 4 \pi}}  \, .
\eea
It is more convenient to drop the Gaunt factor and work with the reduced bispectrum $b_{\ell_1,\ell_2,\ell_3}$ defined as 
\be
B_{\ell_1 \ell_2 \ell_3}^{m_1 m_2  m_3} =  \Gaunt{m_1}{m_2}{m_3}{\ell_1}{\ell_2}{\ell_3} b_{\ell_1 \ell_2 \ell_3} \,.
\ee

Evaluating the 4-fold integral in Eq.~\eqref{full} for an arbitrary template is numerically challenging. This is due to the fact that the three Bessel functions make the integrand highly oscillatory. Moreover the transfer functions  also have oscillatory behaviour. For the CMB bispectrum, for some separable templates such as the local and equilateral shapes, one can change the order of integration over $r$ and $k$ and first perform each individual $k$-integration before calculating the $r$ integral \cite{Liguori:2010hx}. For the galaxy bispectrum however, upon a naive\footnote{As shown in~\cite{DiDio:2015bua}, through a proper treatment, i.e.~by conserving the triangle inequality condition, the integral order can be reversed. Nevertheless this does not help in reducing the dimensionality of the integration.} change of the order of integration, each individual $k$ integral is no longer convergent. Therefore one needs to calculate the 4-dimensional integration numerically which is computationally expensive and inefficient.  It is however possible to simplify this result by performing the $r$ integral analytically. This reduces the dimension of the integral from four to three and gets rid of the three spherical Bessel functions.
We use the results of~\cite{bessel_int}:
 \begin{multline}
\qquad i^{(\ell_1+\ell_2+\ell_3)}\left(\begin{matrix}
    \ell_1       & \ell_2 & \ell_3 \\
  0      & 0 & 0 
\end{matrix}\right)   \int \, dr \ r^2 \, j_{\ell_1}(r k_1) j_{\ell_2}(r k_2) j_{\ell_3}(r k_3)  = \\\frac{\pi \Delta}{4 k_1 k_2 k_3}  \sum_m (-1)^m \sqrt{ \frac{(\ell_1 - m)! (\ell_2 + m)!}{(\ell_1 + m)!(\ell_2-m)!} }  \left(\begin{matrix}
    \ell_1       & \ell_2 & \ell_3 \\
  m     & -m & 0 
\end{matrix}\right) P_{\ell_1}^m (\cos \theta_{13}) P_{\ell_2}^{-m} (\cos \theta_{23})\,,
 \end{multline}
 where $|m| \leq \ell_1, \ell_2 $ and $P^m_\ell$ denote the associated Legendre functions. Therefore, taking $\ell_3$ to be the largest of the $\ell$'s, the sum contains $2\min(\ell_i)+1$ terms. 
Furthermore, $\Delta = 1$ if the $\vk_i$ form a non degenerate triangle, $1/2$ for degenerate triangles and zero if they do not form a triangle. The $\theta_{ij}$ are the angles between the sides $k_i$ and $k_j$ of the triangle.
 
Therefore the reduced bispectrum can be expressed as
 \bea
 b_{\ell_1 \ell_2 \ell_3} &=&\frac{ \left( - 1\right)^{\frac{\ell_1+\ell_2 +\ell_3}{2}}}{\left(\begin{matrix}
    \ell_1       & \ell_2 & \ell_3 \\
  0      & 0 & 0 
\end{matrix}\right)} \left( \frac{2}{\pi} \right)^3 \frac{\pi \Delta}{4}   \fnl^{\rm shape} \sum_m \left( -1 \right)^m 
 \sqrt{ \frac{(\ell_1 - m)! (\ell_2 + m)!}{(\ell_1 + m)!(\ell_2-m)!} }  \left(\begin{matrix}
    \ell_1       & \ell_2 & \ell_3 \\
  m     & -m & 0 
\end{matrix}\right) 
\nonumber \\
&&\times  \int  \d k_1  \d k_2  \d k_3     \prod_{p=1}^3 \left\lbrace  k_p   \Delta_{\ell_p}(k_p,z_p) \right\rbrace   P_{\ell_1}^m (\cos \theta_{13}) P_{\ell_2}^{-m} (\cos \theta_{23}) F_{\rm shape}(k_1, k_2, k_3)\,. \nonumber \\  \label{2.20}
 \eea 
We compute the 3-dimensional integral (\ref{2.20}) using the Suave\footnote{\url{http://www.feynarts.de/cuba/}} Monte Carlo integrator. The transfer functions are computed with the {\sc class}\footnote{\url{http://class-code.net}} code~\cite{Blas:2011rf}, with the LSS implementation described in~\cite{DiDio:2013bqa}.

%%%%%%%%%%%%%%%%%%%%%%%%%%%%%%%%%%%%%%%%
\subsection{Shape correlations in $z\ell m$-space}
Similar to what is done for the CMB \cite{Babich:2004gb, Fergusson:2008ra}, to determine whether two 
non-Gaussian shapes can be distinguished from LSS observations, we can calculate the shape correlations between two theoretically calculated LSS bispectra weighted by the expected signal-to-noise as in a Fisher matrix analysis. For a fixed redshift, the correlation angle (or overlap) between the two angle-averaged spectra in $\ell$-space is defined by
\be
\label{eq:cos}
\cos(B,B') = \frac{B \cdot B' }{\sqrt{B \cdot B}\sqrt{B' \cdot B'}}\,,
\ee
where the scalar product is defined as 
\be \label{eq:scalar_product}
B \cdot B' = \sum_{\ell_1 \le \ell_2 \le \ell_3} \frac{B_{\ell_1 \ell_2 \ell_3} ~  B'_{\ell_1 \ell_2 \ell_3}}{ f_{\ell_1 \ell_2 \ell_3}C_{\ell_1}C_{\ell_2}C_{\ell_3}}.
\ee
We have also introduced the angle-averaged bispectrum 
\be
B_{\ell_1,\ell_2,\ell_3} = \sum_{m_i} \left( \begin{array}{ccc} \ell_1 & \ell_2 & \ell_3 \\ m_1 & m_2 & m_3 \end{array} \right)\left< a_{\ell_1 m_1}^{\Delta_g}(z_1) a^{\Delta_g}_{\ell_2 m_2}(z_2) a^{\Delta_g}_{\ell_3 m_3}(z_3) \right> ,  
\ee
which is simply related to the reduced bispectrum as
\be
B_{\ell_1,\ell_2,\ell_3} = \left( \begin{array}{ccc} \ell_1 & \ell_2 & \ell_3 \\ 0 & 0 & 0 \end{array} \right)  \sqrt{\frac{\left( 2 \ell_1+1 \right) \left( 2 \ell_2+1 \right) \left( 2 \ell_3+1 \right)}{ 4 \pi}} \  b_{\ell_1 \ell_2 \ell_3}\, .
\ee

Furthermore, in Eq.~(\ref{eq:scalar_product}) the sum is taken between $\ell = 3$ and $\ell_{\rm max}$, since the angular bispectra involving $\ell=0,1,2$ depend on perturbations evaluated at the observer position as shown in~\cite{DiDio:2014lka}. The factor $f_{\ell_1,\ell_2,\ell_3}= 1,2,6$ when all the $\ell$'s are different, two of them are the same and all of them are the same, respectively.  Finally, the angular power spectra include the full number counts $C_\ell  = C_\ell^{\Delta_g \Delta_g}$, and is defined as
\be \label{cl_def}
C_\ell^{\Delta_g \Delta_g} = \frac{2}{\pi} \int dk k^2 P_\zeta \left( k \right) \Delta_\ell \left( k \right)  \Delta_\ell \left( k \right) \,,
\ee
where $\Delta_\ell(k)$ are the transfer functions defined in  Eq.~\eqref{numbercounts_transfer} (see Appendix A.4 of~\cite{DiDio:2013bqa} for more details). We have dropped the explicit redshift-dependence of $C_\ell$ and $\Delta_\ell $ as for the rest of our discussion we fix the redshift to be $z=0.55$, apart when specified.

\begin{figure}[htbp]
\begin{center}
\includegraphics[width= .7\textwidth]{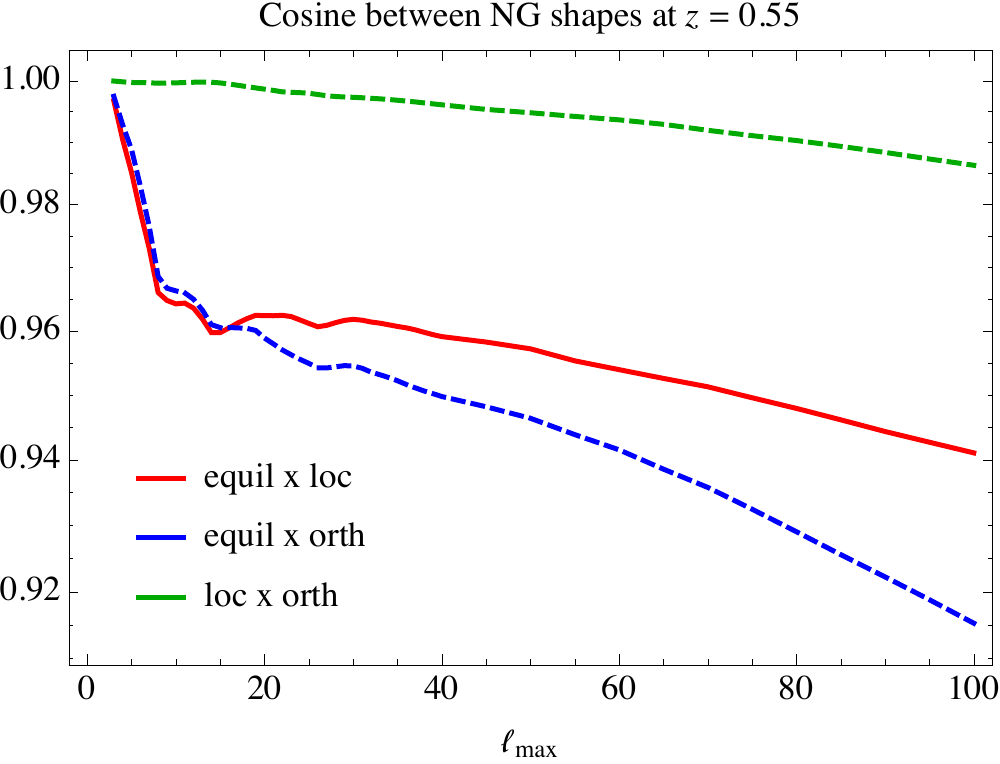}
\caption{Cosine defined through Eq.~\eqref{eq:cos} between the three PNG shapes for galaxy bispectra at $z=0.55$ in $z \ell m$ space as a function the maximum $\ell_\text{max}$ included in the sum. Dashed lines denote negative values.}
\label{fig:cos}
\end{center}
\end{figure}

In Figure~\ref{fig:cos} we plot the cosine of the correlation angle between different shapes in $\ell$-space,
this is evaluated for $\ell $ up to $\ell _{\rm max}=100$. 
As an example, in the case of the local and equilateral shapes the two templates are highly correlated in $\ell$-space with a cosine of about 0.94 for $\ell _{\rm max}=100$. This implies that terms which have a large local component will `leak' into the equilateral shape and give a substantial contribution to $\fnleq$ and vice versa.  Also note that the cosine between the two shapes is slowly reduced as one considers smaller scales. This is due to fact that adding more modes and hence more information helps discriminate the two shapes better. Note also that the orthogonal shape is nearly perfectly anti-correlated with the local shape.
This situation certainly will improve when going to higher $\ell$'s and when considering different redshift since 
$z\ell m$-space contains in principle the same information as Fourier space. 
By comparing with the imprint of PNG on CMB anisotropies, see e.g.~\cite{Lacasa:2011ej}, we remark that different primordial non-Gaussian shapes are much more correlated for LSS observables. While in both cases Fourier scales are mixed by projecting the perturbations on spherical harmonics, for the CMB bispectrum Limber approximation \cite{Limber:1954zz,LoVerde:2008re} can be adopted providing a one to one relation between Fourier and angular scales ($k\simeq (\ell+1/2)/r$). Due to the scaling with power of the matter transfer function the $k$-integrals for LSS are not dominated by the spherical Bessel peaks, but they are getting relevant contributions from several Fourier scales, especially for $(\ell+1/2)/r(z)<k_{\rm eq}$ the integrand  peaks not at $(\ell+1/2)/r$ but rather at $k_{\rm eq}$, the equality scale, where also the matter transfer function peaks. As a consequence, LSS bispectrum is described in terms of a non-trivial 4-dimensional integral even for separable PNG shapes, leading to a large correlation between PNG shapes.

\begin{figure}[ht!]
\begin{center}
\includegraphics[width=0.49 \textwidth]{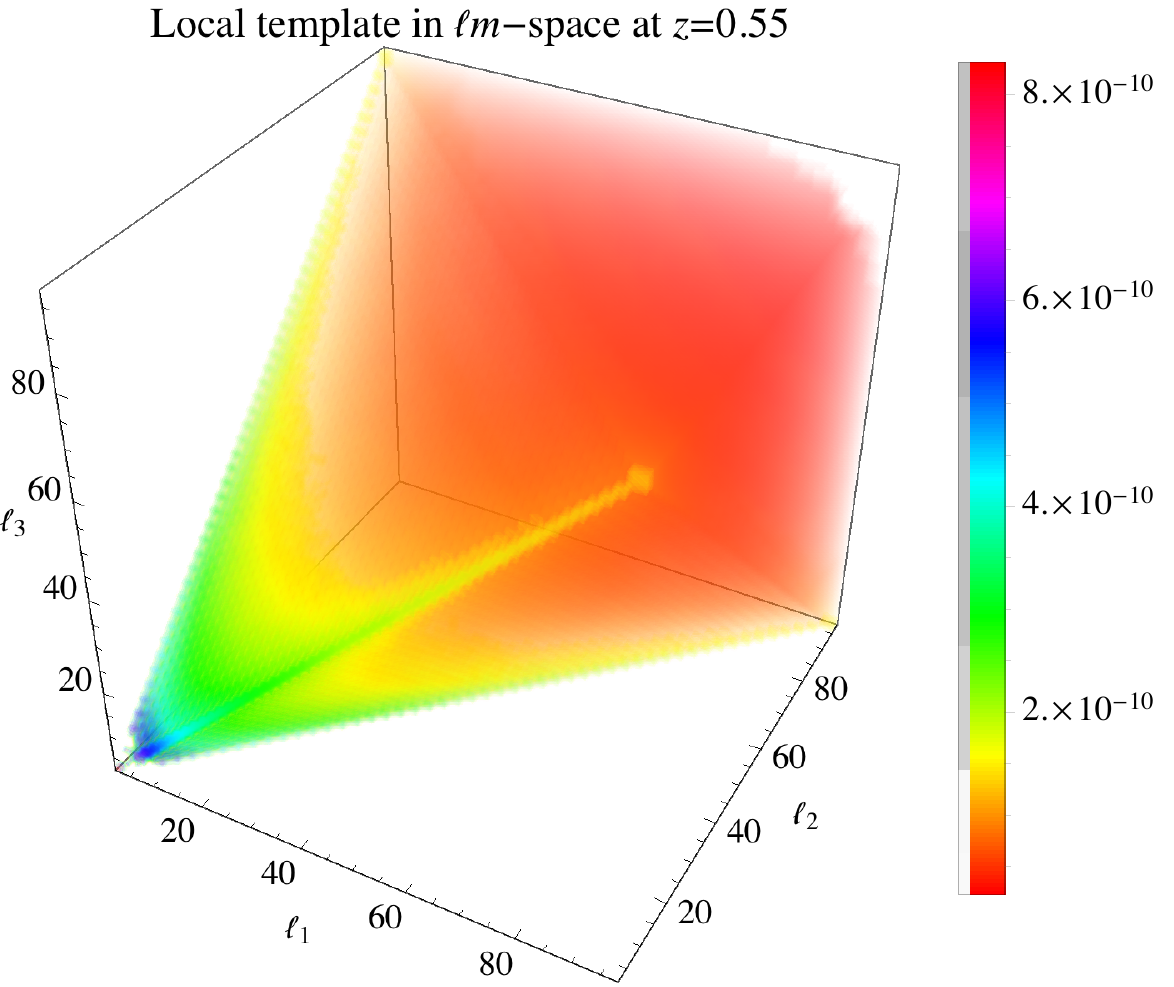} 
\includegraphics[width=0.49 \textwidth]{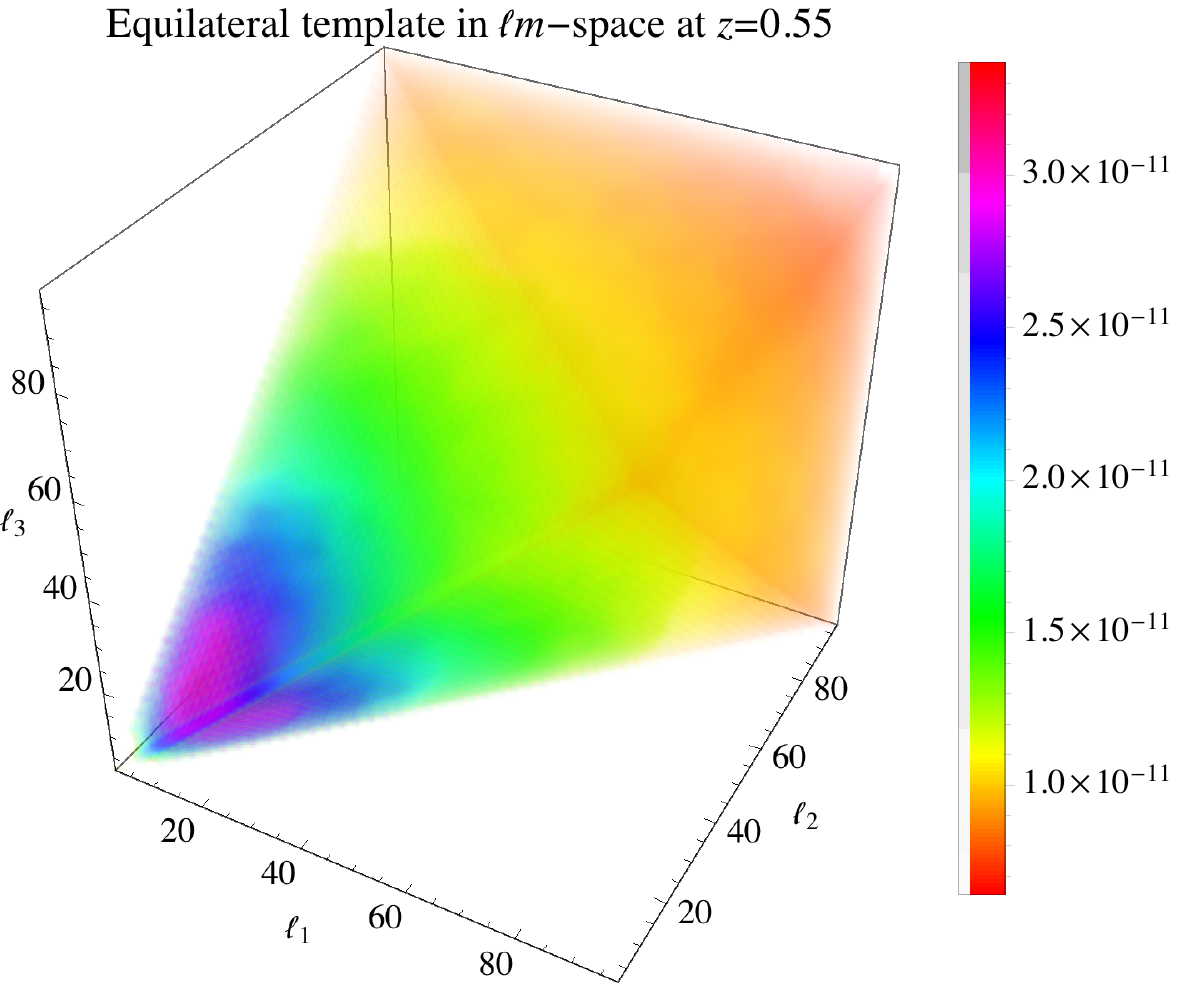}
\\
\includegraphics[width=0.49 \textwidth]{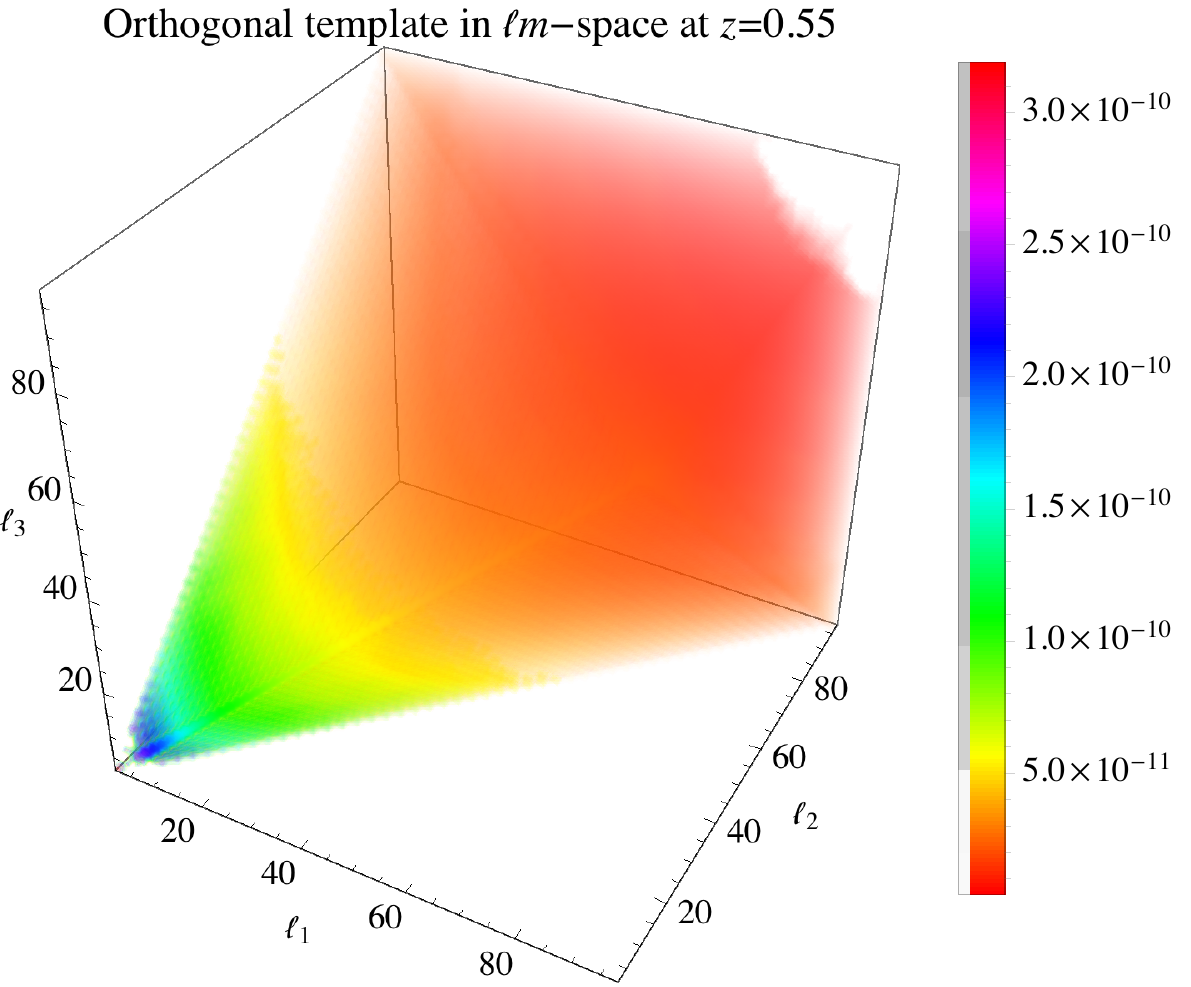}
\caption{We plot the (absolute) value of the bispectrum for the non-Gaussianity templates: local (top left), equilateral (top right) and orthogonal (bottom). We consider three galaxies at the same redshift $z=0.55$
}
\label{fig:shapes}
\end{center}
\end{figure}

In Fig.~\ref{fig:shapes} we present the non-Gaussian shapes in $\ell$-space for local, equilateral and orthogonal templates. The galaxies are taken to be at the same redshift $z=0.55$. As expected the local non-Gaussianity template peaks on the squeezed limit ($\ell_1 \ll \ell_2 \sim \ell_3 $), while the equilateral shape is maximum when $\ell_1$, $\ell_2$ and $\ell_3$ are roughly equal. It should be noted that for the local shape, if only considering the squeezed triangles, the bispectrum signal grows as a function of $\ell_{\rm max}$ while for the equilateral and orthogonal shapes (depending on the squeezing factor) it saturates quickly at $\ell_{\rm max} \simeq 50$.

\subsection{Effective $f_{\rm NL}$ due to projection effects}
\label{efffnl}
Having computed the non-Gaussian templates, we can estimate the systematic shift on the measured amplitude of the primordial non-Gaussianity induced by relativistic projection effects. This is done by projecting them on the relevant shape (see appendix~\ref{app:systematic} for a short review):
\be \label{eq:fnlshape}
\fnl^\text{shape} = \frac{\sum_{\ell_1 \le \ell_2 \le \ell_3}~ \frac{B_{\ell_1 \ell_2 \ell_3} ~  B^\text{shape}_{\ell_1 \ell_2 \ell_3}}{C_{\ell_1}C_{\ell_2}C_{\ell_3}}}
{\sum_{ \ell_1 \le \ell_2 \le \ell_3 }  ~ \frac{ \left( B^\text{shape}_{\ell_1 \ell_2 \ell_3} \right)^2}{C_{\ell_1}C_{\ell_2}C_{\ell_3}}\
} \, .
\ee
As we discuss in the next section, for the estimate of  $f_{\rm NL}$ using the bispectrum calculated from the consistency relation, we only sum over squeezed configurations for which $10 \ \ell_1 \leq \min(\ell_2,\ell_3)$. This is because the consistency relation bispectrum is valid only in the squeezed limit. On the other hand, to estimate $f_{\rm NL}$ from the bispectrum calculated using the second-order perturbation theory, since it is valid for all configurations, we present our results both when summing  only over squeezed configurations and over all configurations involving $\ell$'s up to $\ell_{max}$. 

{We believe that this is the definition of $f_{\rm NL}$ for a given shape as it can be used in an observation. Note that it is important, especially when not projecting onto a given shape,  to divide the bispectrum by the \emph{observed} power spectrum and not by some theoretical power spectrum.  This may explain the significantly larger results obtained in~\cite{Umeh:2016nuh}. In this reference,  the effective  $f_{\rm NL}$ is defined as the amplitude of the bispectrum normalized with respect to  the \emph{Newtonian} power spectrum squared.  Since the relativistic  bispectrum  has much more power at large scales, when normalizing with respect to the Newtonian power spectrum the  effective non-Gaussianity  is seen to be large.
}
 
%%%%%%%%%%%%%%%%%%%%%%%%%%%%%%%%%

\section{Contributions to the bispectrum from relativistic corrections}\label{secgrcorr}
In this section we summarize the bispectrum generated by  projection effects as obtained 
from the second-order calculation of \cite{DiDio:2015bua} and
from the consistency relation \cite{Kehagias:2015tda}. For the former, as mentioned in the introduction, we only consider a subset of terms. We do not go into the details of how the bispectrum is calculated in each case and we simply report the final results. We refer the reader to corresponding papers for their derivations. 

%%%%%%%%%%%%%%%%%%%%%%%%%%%%%

\subsection{Bispectrum from second-order perturbation theory}
\label{bisperturb}
The second order relativistic calculation of the observed galaxy number counts has been developed by three independent groups~\cite{Yoo:2014sfa,Bertacca:2014dra,DiDio:2014lka} and recently compared partially in~\cite{Nielsen:2016ldx}. 
In particular, in \cite{DiDio:2014lka} the second-order galaxy number counts are calculated using a geometrical approach based on the so-called geodesic light-cone coordinates~\cite{Gasperini:2011us}, 
and it is valid for any metric theory with perturbations around FLRW metric. 
However, one has to make the assumption of GR, when using the dynamical equations to relate the intrinsically second-order quantities such as $\delta^{(2)}$ to the gravitational potential. 

The second order number counts have been used~\cite{DiDio:2015bua} to derive the tree-level bispectrum for the leading terms on intermediate to large redshift and sub-Hubble scales. For Gaussian initial conditions, the leading contributions to the 3-point function are in the form of 
\be
\left \langle \Delta_g^{(2)}\Delta_g^{(1)} \Delta_g^{(1)}\right \rangle_c +{\rm perms.} \,,
\ee
where $\Delta_g^{(1)}$ and $\Delta_g^{(2)}$ are the linear and second-order galaxy number counts. In the full second-order result, $\Delta^{(2)}_g$ contains hundreds of terms. Here we only consider a subset of the contributions and calculate the effective $f_{\rm NL}$ due to these terms which can be misinterpreted as local, equilateral or orthogonal PNG. One can organize the contributions to $\Delta_g^{(2)}$ in terms of the number of spatial derivatives acting on the Bardeen potentials. In our analysis we only consider a subset of the three- 
and four-derivative terms with respect to the Bardeen potentials.  
For the linear galaxy number counts $\Delta^{(1)}_g$ we only include the Newtonian terms, i.e. the density and redshift-space distortion 
\be \label{density_and_rsd}
\Delta^{(1)}_g =  \delta + \HH^{-1} \partial_r^2 v,  
\ee 
where for sake of simplicity we omit the label $^{(1)}$, therefore all the perturbations have to be interpreted at first order unless they are explicitly denoted as second order. Among the second-order contributions to galaxy number counts, we consider the subset of three- 
and four-derivative terms defined as
\be
\tilde{\Delta}^{(2)}_g = \Delta^{(2)}_{g,3dA} + \Delta^{(2)}_{g,3dB} + \Delta^{(2)}_{g, \rm N \times \rm L}\,.
\ee
The three-derivative terms are
\bea \label{3diff} 
\Delta^{(2)}_{g,3dA} &=&  
\left( \frac{2}{\deta\HH} +\frac{\dot\HH}{\HH^2} \right)
 \delta  \ \partial_r v
+\frac{1}{\HH}\left(1+\frac{4}{ \deta\Hcal }+3 \frac{\dot\HH}{\HH^2  }\right)\partial_r v \partial_r^2 v 
-  \frac{1}{\HH} \dot\delta \partial_r v  \, ,  \\
\Delta^{(2)}_{g,3dB} &=&  
\frac{1}{\HH^2} \left[  
\Phi_W\partial^3_r  v -\partial_r^2 \Phi_W  \partial_r v\right]+\frac{1}{\HH}  \Phi_W \partial_r \delta \, .
\, \label{e:SIS-3der-B}
\eea
where we have introduced the Weyl potential
\be
\Phi_W = \frac{\Psi + \Phi}{2} \,,
\ee
and classified the terms such that only $\Delta^{(2)}_{g,3dB}$ involves contribution from the metric perturbations. 

Among the four-derivative terms, we consider the Newtonian $\times$ lensing terms given by
\be\label{newtXlens}
\Delta^{(2)}_{g, \rm N \times \rm L} = -2 \delta \kappa + \nabla_a\delta \nabla^a \psi+\HH^{-1} \left[-2 \kappa \partial^2_r v   + \nabla_a \left(\partial^2_r v\right)\nabla^a\psi\right],
\ee
where $\nabla_a$ denotes the covariant derivative on the 2-sphere, we have introduced the lensing potential
\be
\psi = -2 \int_0^r dr' \frac{r - r'}{rr'} \Phi_W  \,,
\ee
and the first order magnification or convergence 
\be
\kappa = -\frac{\Delta_\Omega \psi}{2} \, .
\ee
The bispectrum including terms of the form
\be \label{eq:NLdeltadelta}
\langle \Delta^{(2)}_{g, \rm N \times \rm L} \left( \bn_1 , z_1 \right) \delta\left( \bn_2 , z_2 \right)  \delta \left( \bn_3 , z_3 \right)\rangle 
\ee
has been computed in~\cite{DiDio:2015bua}. Here we do not repeat the calculation and we just report the final results for each single term of Eq.~\eqref{newtXlens}:
\bea
b^{\kappa\delta}_{\ell_1 \ell_2 \ell_3}\left(z_1,z_2,z_3\right)&=& \left[ c_{\ell_2}^{\ka \delta}( z_1, z_2) c_{\ell_3}^{\delta \delta}( z_1 , z_3)  +c_{\ell_3}^{\ka \delta}( z_1, z_3) c_{\ell_2}^{\delta \delta}( z_1, z_2) \right]  + \mbox{perms.} \,, \\
b^{\nabla\delta\nabla\psi}_{\ell_1 \ell_2 \ell_3}\left(z_1,z_2,z_3\right)&=&A_{\ell_1 \ell_2 \ell_3} \left[\sqrt{\frac{\ell_3 \left( \ell_3 + 1 \right)}{\ell_2 \left( \ell_2 + 1 \right)}} c_{\ell_2}^{\ka \delta} \left( z_1, z_2\right) c_{\ell_3}^{\delta \delta} \left( z_1 , z_3 \right)
 \right. \nonumber \\  && \qquad   \left.
~+~ \sqrt{\frac{\ell_2 \left( \ell_2 + 1 \right)}{\ell_3 \left( \ell_3 + 1 \right)}} c_{\ell_3}^{\ka \delta} \left( z_1, z_3\right) c_{\ell_2}^{\delta \delta} \left( z_1 , z_2 \right) \right]
+ \mbox{perms.}\,, \label{e3:nabla-d-nabla-psi} 
\\
b_{\ell_1 \ell_2 \ell_3}^{ v' \kappa}\left(z_1,z_2,z_3\right)&=& \left[
c_{\ell_2}^{v' \delta} \left( z_1, z_2 \right)c_{\ell_3}^{\ka  \delta} \left( z_1, z_3 \right)
+ c_{\ell_2}^{\ka  \delta} \left( z_1, z_2 \right)c_{\ell_3}^{v' \delta} \left( z_1, z_3 \right)
\right] +  \mbox{perms.}\,, \label{e3:v'k}
\\
b^{\nabla v'\nabla\psi}_{\ell_1 \ell_2 \ell_3}\left(z_1,z_2,z_3\right) &=&
A_{\ell_1 \ell_2 \ell_3} \left[\sqrt{\frac{\ell_3 \left( \ell_3 + 1 \right)}{\ell_2 \left( \ell_2 + 1 \right)}} c_{\ell_2}^{\ka \delta} \left( z_1, z_2\right) c_{\ell_3}^{v' \delta} \left( z_1 , z_3 \right) \right. \nonumber \\  && \qquad   \left.
~ + ~  \sqrt{\frac{\ell_2 \left( \ell_2 + 1 \right)}{\ell_3 \left( \ell_3 + 1 \right)}} c_{\ell_3}^{\ka \delta} \left( z_1, z_3\right) c_{\ell_2}^{v' \delta} \left( z_1 , z_2 \right) \right]
+ \mbox{perms.}\,, \label{e3:nabla-v-nabla-psi}
\eea
where we have defined
\bea\label{def:Afactor}
A_{\ell_1 \ell_2 \ell_3}\equiv\frac{1}{2}\frac{\left[
\left( \begin{array}{ccc} \ell_1 & \ell_2 & \ell_3 \\ 0 & 1 & -1 \end{array} \right) +  \left( \begin{array}{ccc} \ell_1 & \ell_2 & \ell_3 \\ 0 & -1 & 1 \end{array} \right)
\right] }{\left( \begin{array}{ccc} \ell_1 & \ell_2 & \ell_3 \\ 0 & 0 & 0 \end{array} \right)} \, .
\eea
The angular spectra are defined following the convention of Eq.~\eqref{cl_def}, where the transfer function associated are
\bea
\De^\delta_{\ell}(z,k) &=&  T_\delta\left(\eta(z) , k \right)j_\ell\left(kr(z)\right)\,, 
\\
\Delta_\ell^\ka(z,k) &=& \ell(\ell+1)\int_0^{r(z)}dr\frac{r(z)-r}{r(z)r}T_{\Psi+\Phi}(\eta_0-r,k)j_\ell(kr)\, , 
\\
\Delta^{v'}_{\ell}(z,k)&=&\frac{k}{\HH(z)}T_v(k,\eta)j_\ell''(kr) \, .
\eea
To generalize Eq.~\eqref{eq:NLdeltadelta} from the linear $\delta$ to the number counts $\Delta^{(1)}$, as considered in Eq.~\eqref{density_and_rsd}, we simply replace $\De^\delta_{\ell}(z,k)$ with $\De^\delta_{\ell}(z,k)+\De^{v'}_{\ell}(z,k)$

For the three-derivative terms defined in Eqs.~(\ref{3diff}) and (\ref{e:SIS-3der-B}) we can proceed as above (see Appendix~\ref{sec:3diff_appendix} for details).
Namely, we compute the bispectra
\be
\label{eq:bispectra_3d}
\langle \Delta^{(2)}_{g,3dA} \left( \bn_1, z_1 \right) \delta \left( \bn_2 , z_2 \right) \delta \left( \bn_3 , z_3 \right) \rangle \quad \text{and} \quad \langle\Delta^{(2)}_{g,3dB} \left( \bn_1, z_1 \right) \delta \left( \bn_2 , z_2 \right) \delta \left( \bn_3 , z_3 \right) \rangle \,,
\ee
then the full results can be obtained by simply replacing $\De^\delta_{\ell}(z,k)$ with $\De^\delta_{\ell}(z,k)+\De^{v'}_{\ell}(z,k)$. 
The bispectrum involving $\Delta^{(2)}_{g,3dA}$ can then be written as the sum of the following terms
\bea
b^{\delta v}_{\ell_1 \ell_2 \ell_3} \left( z_1 , z_2 , z_3 \right)&= & \left( \frac{2}{r \HH} + \frac{\dot \HH}{\HH^2} \right)_{\!\! z=z_1} \!\!\!\!\!\!\!\left[C^{\delta \delta}_{\ell_2} \left( z_1 , z_2 \right) C^{v \delta}_{\ell_3} \left( z_1 , z_3 \right) 
+C^{\delta \delta}_{\ell_3} \left( z_1 , z_3 \right) C^{v \delta}_{\ell_2} \left( z_1 , z_2 \right)
 \right] 
 \nonumber \\
 &&
 + \mbox{perms.}\, , \\
 %%%%%%%%%%%
 b^{v' v}_{\ell_1 \ell_2 \ell_3} \left( z_1 , z_2 , z_3 \right)&= & \left(1+\frac{4}{ \deta\Hcal }+3 \frac{\dot\HH}{\HH^2  } \right)_{\!\! z=z_1} \!\!\!\!\!\!\!
\left[C^{v' \delta}_{\ell_2} \left( z_1 , z_2 \right) C^{v \delta}_{\ell_3} \left( z_1 , z_3 \right) 
+C^{v' \delta}_{\ell_3} \left( z_1 , z_3 \right) C^{v \delta}_{\ell_2} \left( z_1 , z_2 \right)
 \right] 
 \nonumber \\
 &&
 + \mbox{perms.}\, ,
 %%%%%%%%
 \\
  b^{\dot\delta v}_{\ell_1 \ell_2 \ell_3} \left( z_1 , z_2 , z_3 \right)&= & -
\left[C^{\dot \delta \delta}_{\ell_2} \left( z_1 , z_2 \right) C^{v \delta}_{\ell_3} \left( z_1 , z_3 \right) 
+C^{\dot\delta \delta}_{\ell_3} \left( z_1 , z_3 \right) C^{v \delta}_{\ell_2} \left( z_1 , z_2 \right)
 \right] +  \mbox{perms.}\, .
\eea
Similarly, for the bispectrum involving $\Delta^{(2)}_{g,3dB}$ we have
\bea
 b^{\Phi v''}_{\ell_1 \ell_2 \ell_3} \left( z_1 , z_2 , z_3 \right)&= &
 \left[C^{\Phi \delta}_{\ell_2} \left( z_1 , z_2 \right) C^{v'' \delta}_{\ell_3} \left( z_1 , z_3 \right) 
+C^{\Phi \delta}_{\ell_3} \left( z_1 , z_3 \right) C^{v'' \delta}_{\ell_2} \left( z_1 , z_2 \right)
 \right]
 \!\!+\!  \mbox{perms.}\,,
 \\
  b^{\Phi'' v}_{\ell_1 \ell_2 \ell_3} \left( z_1 , z_2 , z_3 \right)&= & - 
 \left[C^{\Phi'' \delta}_{\ell_2} \left( z_1 , z_2 \right) C^{v \delta}_{\ell_3} \left( z_1 , z_3 \right) 
+C^{\Phi'' \delta}_{\ell_3} \left( z_1 , z_3 \right) C^{v \delta}_{\ell_2} \left( z_1 , z_2 \right)
 \right]
 \!\!+\!  \mbox{perms.}, \qquad
  \\
  b^{\Phi \delta'}_{\ell_1 \ell_2 \ell_3} \left( z_1 , z_2 , z_3 \right)&= & - 
 \left[C^{\Phi \delta}_{\ell_2} \left( z_1 , z_2 \right) C^{\delta' \delta}_{\ell_3} \left( z_1 , z_3 \right) 
+C^{\Phi \delta}_{\ell_3} \left( z_1 , z_3 \right) C^{\delta \delta}_{\ell_2} \left( z_1 , z_2 \right)
 \right]
\!\! +\!  \mbox{perms.}\,,
\eea 
To compute the bispectra~\eqref{eq:bispectra_3d}, we have introduced the following angular transfer functions
\bea\label{deltadot_trans}
\De^{\dot\delta}_{\ell}(z,k) &=& \frac{1}{\HH(z)} \left( \partial_{\eta} T_{\delta}\left( k , \eta \right)\right) j_\ell \left( k r \right)
\, , \\
\De^{\delta'}_{\ell}(z,k) &=&\frac{k}{\HH(z)}  T_\delta(k,\eta)j_\ell'(kr)
\, , \\
\De^{v''}_{\ell}(z,k) &=&\left(\frac{k}{\HH(z)} \right)^2 T_v(k,\eta)j_\ell'''(kr)
\, , \\
\De^{\Phi}_{\ell}(z,k) &=& \frac{1}{2} T_{\Psi +\Phi} \left( k , \eta \right) j_\ell (kr)
\, , \\
\De^{\Phi''}_{\ell}(z,k) &=&\frac{1}{2} \left(\frac{k}{\HH(z)} \right)^2  T_{\Psi +\Phi} \left( k , \eta \right) j''_\ell (kr)
\, .
\eea
We remark that, while the three-derivative terms, i.e. $\Delta^{(2)}_{g,3dA}$ and $\Delta^{(2)}_{g,3dB}$, are suppressed by single spatial derivative with respect to the leading terms, the amplitudes of the bispectra in Eq.~\eqref{eq:bispectra_3d} are further reduced since they involve always a power spectrum with Bessel functions out of phase. Therefore, as already pointed out for the leading terms in~\cite{DiDio:2015bua}, these bispectra are highly suppressed for galaxies at the same redshift, while for galaxies at different redshifts we recover the suppression expected by the derivative counting scheme. It would be interesting to do the full 3D analysis and consider the bispectra at several redshifts.

%%%%%%%%%%%%%%%%%%%%
\subsection{The squeezed-limit bispectrum from consistency relation}\label{grconsis}
In \cite{Kehagias:2015tda}, Kehagias {\it et al.} derived a non-perturbative expression for the observed galaxy bispectrum due to projection effects. Following the approach of \cite{Peloso:2013zw, Kehagias:2013yd, Creminelli:2013mca}, the squeezed limit bispectrum was  obtained by correlating the observed galaxy power spectrum in the presence of a long mode with the observed galaxy overdensity on that large wavelength. This bispectrum is exactly computable since the effect of a long mode up to  a gradient on the short-scale dynamics is simply a change of frame (or a residual gauge transformation).

Reporting only their final result, the angular reduced bispectrum for three galaxies at redshifts $z_1,z_2,z_3$ in the squeezed limit is given by
\bea\label{fullconsist}
\lim_{\ell_1\ll \ell_2,\ell_3}  b_{\ell_1,\ell_2,\ell_3}(z_1,z_2,z_3) &=&  \bigg[C_{\ell_1}^{\Delta_{\rm g} d}(z_1, z_2) + C_{\ell_1}^{\Delta_{\rm g} \Delta z}(z_1, z_2)\frac{\de}{\de z_2}  \nn \\
&&
+\frac{1}{2}\left( \ell_1(\ell_1 + 1)  - \ell_2(\ell_2 + 1) 
+\ell_3(\ell_3 + 1)\right)C^{\Delta_{\rm g} I}_{\ell_1}(z_1, z_2) \bigg]  \nn \\
&&
 \times  C^{\Delta_{\rm g} \Delta_{\rm g}}_{\ell_3}(z_2, z_3)  + (2 \leftrightarrow 3).
\eea
The first contribution arises since the presence of the long mode alters the relation between the local and the observed number counts. $d$ is a combination of linear redshift perturbation $\Delta z$, luminosity-distance perturbation $\delta{\mathcal D}_l$ and volume perturbations $\delta V$, generated by the long mode\footnote{We include the evolution and magnification bias parameters, $f_{\rm evo}$ and $s$, for completeness, but we remind the reader that the numerical results are derived for $f_{\rm evo}=0$ and $s=0$.}
\be \label{d_term_definition}
d \equiv (3-f_{\rm evo}) \frac{\Delta z}{1+z} - 5s \ \delta{\mathcal D}_\ell+\delta V.
\ee
The explicit expressions for $\Delta z, \delta {\mathcal D}_\ell$ and $\delta V$ are given in Appendix~\ref{explicit_fullconsist}. The second contribution in the expression for the bispectrum, is purely due to the fact that in the presence of the long mode the observed and background redshifts of the galaxies differ. As explained in Appendix~\ref{explicit_fullconsist}, the redshift derivative can be written in terms of a radial and a time derivative. Of these two, the radial derivative appears in standard calculations of the bispectrum, and in the following we only keep the time derivative. The last term arises since the angle at which the galaxies are observed is modified with respect to their background angular position, and the explicit expression for $I$ is also given in Appendix~\ref{explicit_fullconsist}. Since we are working inside the horizon and for configurations which are not very squeezed, this last term should be handled with care. When evaluated at equal redshifts, its contribution is expected to be proportional to spurious terms that go like the second derivative of the long mode. These should be discarded since there are several other similar terms that the consistency relation cannot capture. We keep it here since at unequal redshifts it does contain a contribution proportional to the gradient of the long mode, but we ignore it in the following sections where everything is evaluated at the same $z$.  Let us stress that in \cite{Kehagias:2015tda} this term was consistently not included in the estimate of $f_{\rm NL}$.

This relation is valid in the squeezed limit, that is when $\ell_1 \ll \ell_2, \ell_3$. The most important feature is that it is valid even when the number counts associated to the short scales $\ell_2, \ell_3$ are in the non-linear regime. 

It captures the terms that scale as\footnote{We denote the galaxy separation in comoving space with $\lambda$. The suffix $L$ refers to a long mode while $S$ to a short mode such that $\lambda_L \gg \lambda_S.$ The same notation applies also for the metric perturbations.} ${\cal H}^2 \lambda_L^2$ and ${\cal H} \lambda_L$  as one takes the squeezed limit of the galaxy bispectrum as we have accounted the effect of the long mode up to a gradient through a coordinate transformation. This is possible as long as the evolution of the Universe is adiabatic (single-field inflation and a subsequent evolution that conserves $\zeta$ on the scales of interest).

In general, we also expect corrections going like the curvature induced by the long-wavelength mode, that is $(\Phi_L \Delta_{\rm g}/ \lambda_L^2\mathcal{H}^2)$. If the long mode is super-Hubble, these corrections are subdominant. However, when the long mode is sub-Hubble - but still much larger than the short scales - these curvature corrections are parametrically larger than relativistic corrections other than
 the lensing and redshift space distortions which behave as $(\Phi_L \Delta_{\rm g}/\lambda_L \lambda_S {\mathcal H}^2)$.

In \cite{Kehagias:2015tda} the estimate of $f_{\rm NL}$ due to projection effects was made by computing the amplitude of the terms in the bispectrum of Eq.~\eqref{fullconsist} which schematically have the form $b \sim \langle (\Phi \Delta_{\rm g}^{(1)}) \Delta_{\rm g}^{(1)} \Delta_{\rm g}^{(1)} \rangle $. The motivation for this simplification was that since the primordial bispectrum of the local shape has this form in Fourier space, one expects a similar scaling also for the number counts bispectrum. 
The idea was that only terms with a similar scaling would give rise to the local-type $f_{\rm NL}$ that could be misinterpreted as the primordial signal. 
However, as we have discussed in the previous section, a more accurate estimate of $f_{\rm NL}$ is obtained by projecting the full bispectrum onto a particular shape, in this case the local shape.  Moreover, we have seen that different shapes have a high overlap in harmonic space which means that even terms not scaling as the local template for number counts will contribute to the effective $f_{\rm NL}$. 
In the next section we revise our previous estimate using this new, more precise approach.
We checked that the simple estimator of \cite{Kehagias:2015tda} gives the result correct within an order of magnitude.

%%%%%%%%%%%%%%%%%%
\section{Results}\label{secresults}

We now estimate the contamination of the primordial non-Gaussianity signal due to several terms appearing in the relativistic calculation of the observed galaxy bispectrum. We will see that these projection effects, if ignored, would induce a systematic error of order $\Delta f_{\rm NL} \simeq {\mathcal O}(1)$ for different shapes. Since upcoming LSS surveys have a target sensitivity of  that order of magnitude for the local shape \cite{Dore:2014cca,Tellarini:2016sgp}, these effects must be computed and taken into account in the modeling of the bispectrum. For the equilateral shape, achieving such a sensitivity level is known to be more challenging \cite{Baldauf:2016sjb}. In this section we present the level of non-Gaussianity generated from a subset of effects for local, equilateral and orthogonal shapes. 
Having developed the tools to do this, as presented in previous sections, our purpose is only to estimate the order of magnitude of these effects. We leave a full detailed calculation for future work.

%%%%%%%%%%%%%%%%%%%
\subsection{Local non-Gaussianity}\label{localng}
\begin{figure}[ht!]
\begin{center}
\includegraphics[width=0.45 \textwidth]{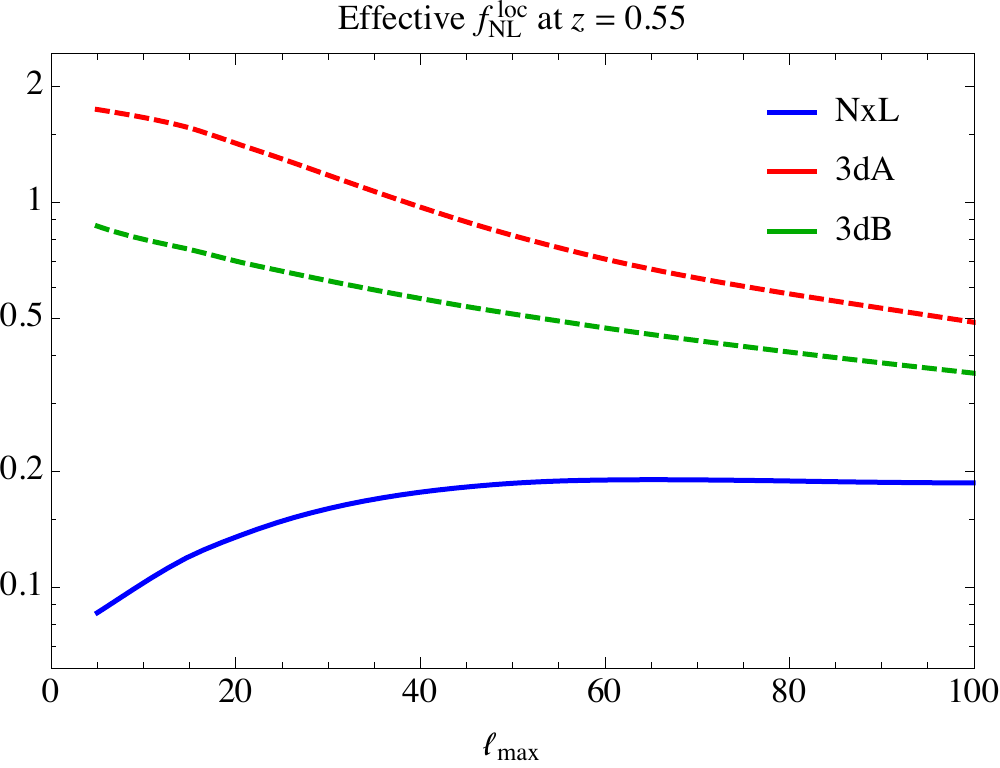} 
\includegraphics[width=0.45 \textwidth]{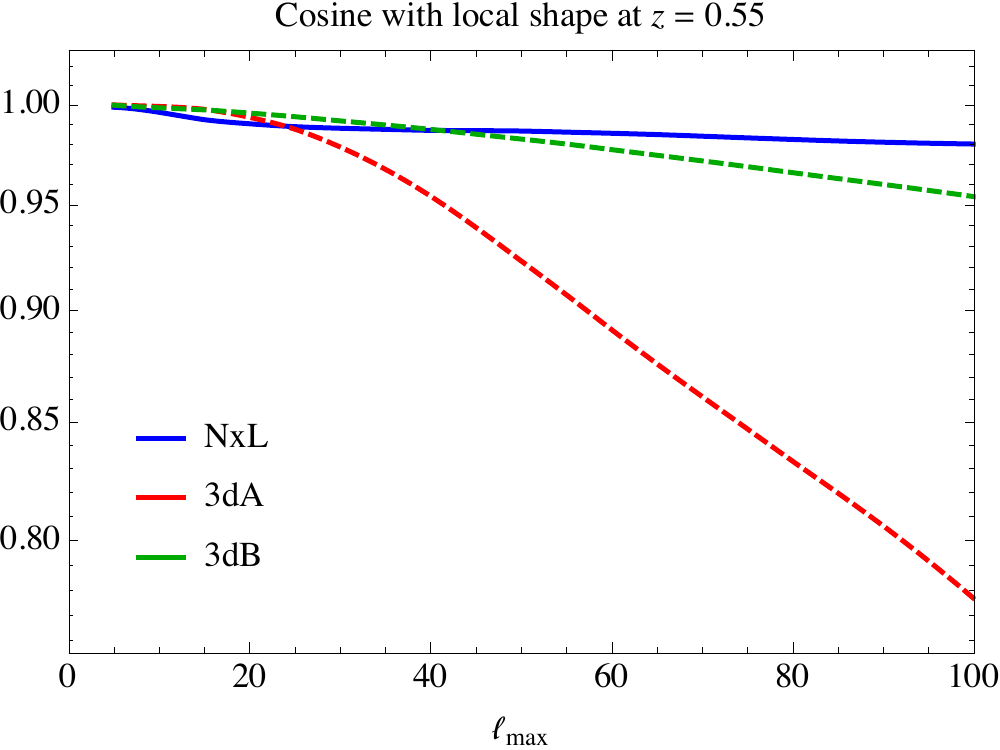}
\includegraphics[width=0.45 \textwidth]{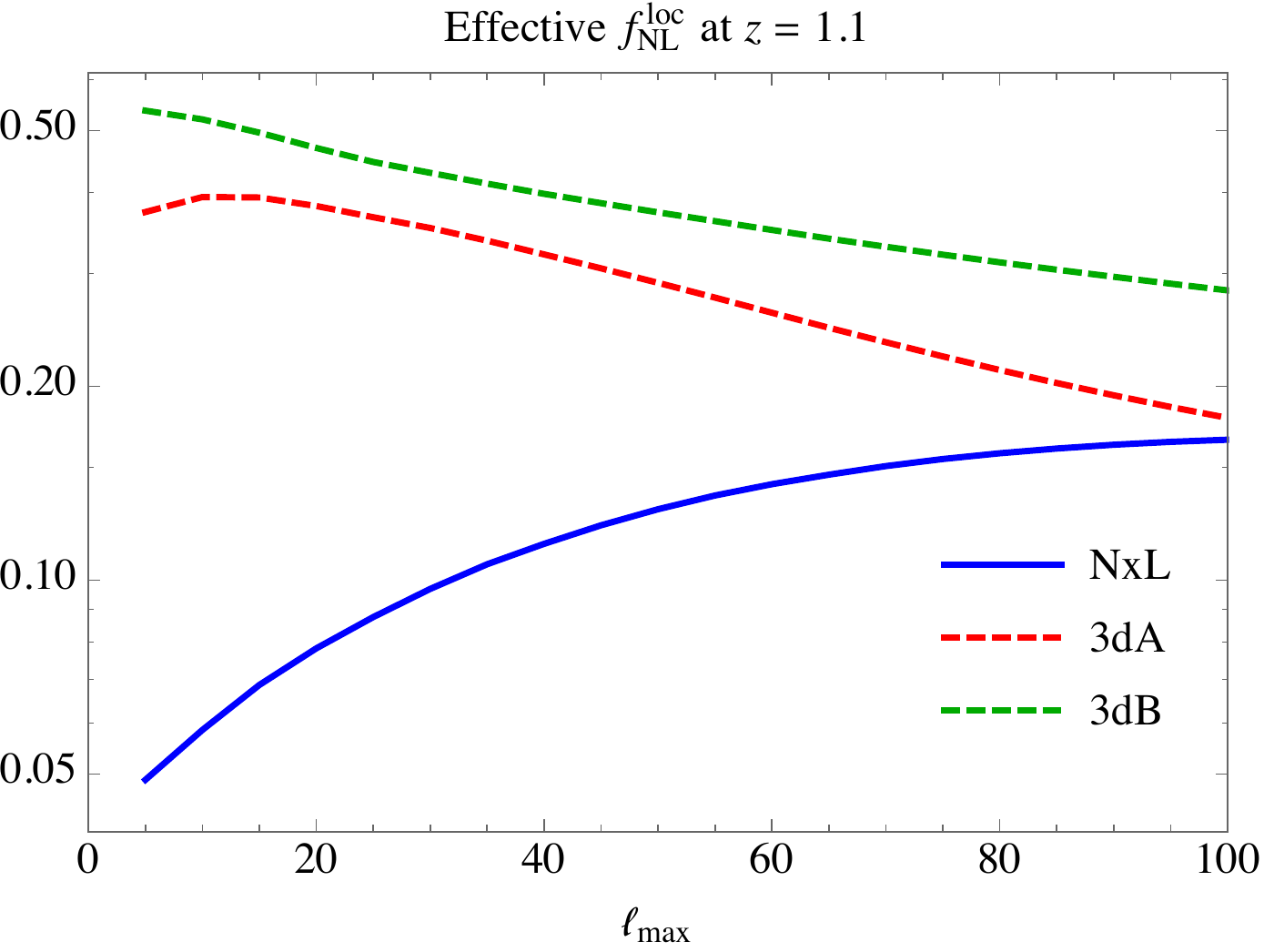} 
\includegraphics[width=0.45 \textwidth]{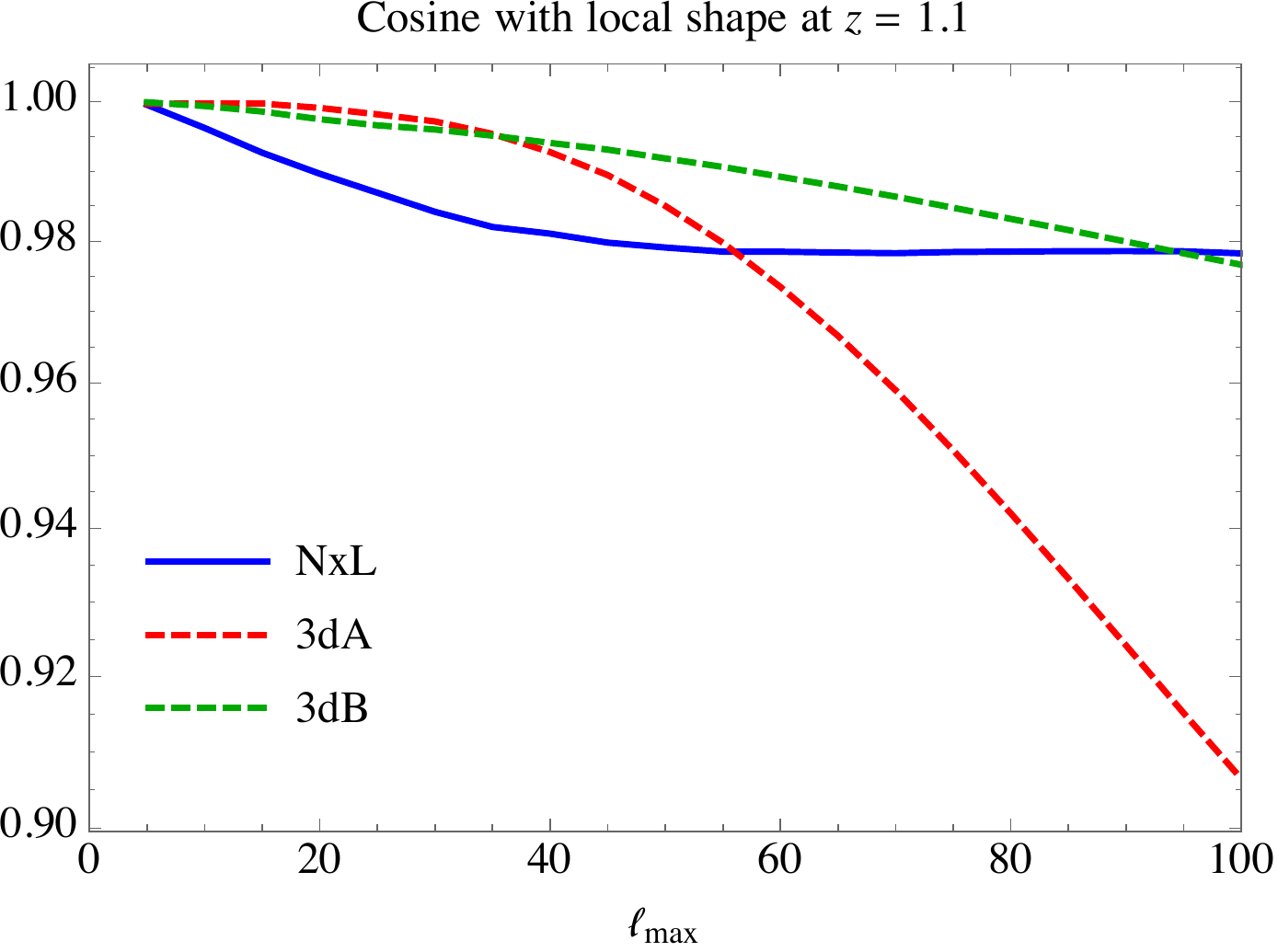}
\caption{Effective local non-Gaussianity (left) and cosine with local shape (right)  from  Newtonian~$\times$~Lensing  and the three-derivative terms summing over all configurations up to $\ell_{max}$ at equal redshifts $z=0.55$ (top) and $z=1.1$ (bottom). Dashed lines indicate negative values.
}
\label{fig:fnllocall}
\end{center}
\end{figure}

We estimate the effective local non-Gaussianity generated by projection effects. We do this by projecting some of the terms appearing in the second-order calculation, namely the  ``Newtonian $\times$ Lensing"  and the three-derivative terms described in section \ref{efffnl}, on the local template as discussed in section \ref{bisperturb}.  We also plot the results from the consistency relation 'CR'.
Results are shown in Fig.~\ref{fig:fnllocall} for two different redshifts ($z=0.55$ and $z=1.1$). In particular, we see that the contamination at $\ell_{max}=100$
and $z=0.55$ is
\begin{align}
&\fnl^{\rm loc, \, N \times L} \simeq 0.19, \\
 &\fnl^{\rm loc, \, 3dA}  \simeq -0.49, \\ 
 &\fnl^{\rm loc, \, 3dB}  \simeq -0.36 \,. 
 \end{align}
We also note that the cosine with the local shape is very close to one for all three, although we would naively not expect this as they have a different scaling than that of the local terms. The cosine of the  `Newtonian $\times$ Lensing" term drops very slowly with $\ell_{max}$, and is thus expected to give a large contamination even when considering smaller scales. 

\begin{figure}[htbp]
\begin{center}
\includegraphics[width=0.45 \textwidth]{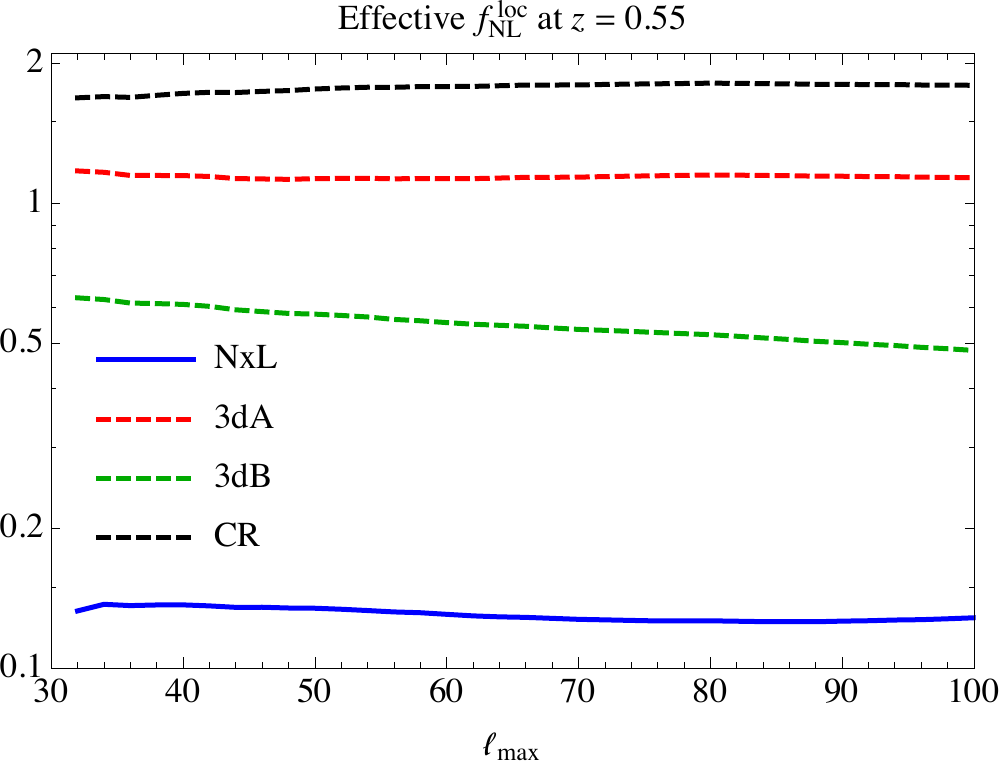} 
\includegraphics[width=0.45 \textwidth]{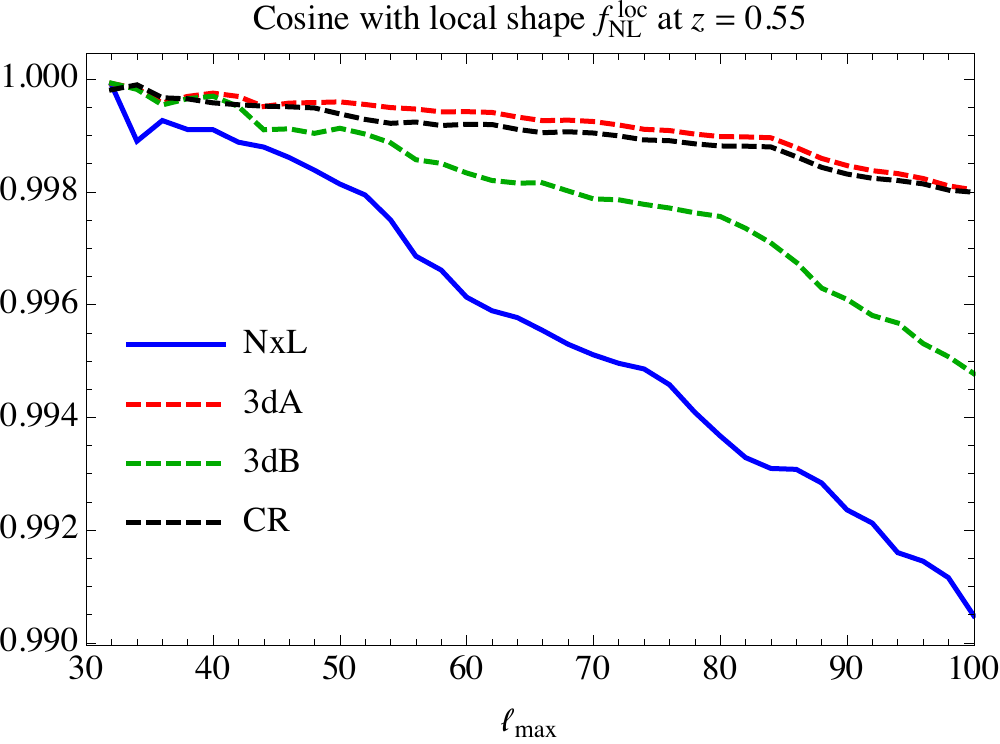}
\caption{Effective local non-Gaussianity (left) and cosine  with local shape (right) from various contributions summing over squeezed configurations only up to $\ell_{max}$ at equal redshift $z=0.55$. Dashed lines indicate negative values. 'CR' stands for consistency relation.}
\label{fig:fnllocsqueezed}
\end{center}
\end{figure}

It is also interesting to consider only squeezed configurations, since in Fourier space the local shape has a characteristic behaviour in that limit that could potentially distinguish it. Figure \ref{fig:fnllocsqueezed} shows the resulting $\fnl$ of the local shape from Newtonian $\times$ Lensing, the three-derivative, and consistency relation terms considering only squeezed configurations, i.e. $ \left| \ell_1 - \ell_2 \right| \le 10 \ell_3 \le  \ell_1 +\ell_2$ , as a function of the non-linear cutoff $\ell_\text{max}$.
We find that they induce a contamination at $\ell_{max}=100$ and $z=0.55$ of
\begin{align}
&\fnl^{\rm loc, \, N\times L}  \simeq 0.13, \\ 
&\fnl^{\rm loc, \, 3dA}  \simeq -1.14, \\
&\fnl^{\rm loc, \, 3dB}  \simeq -0.48\, , \\
& \fnl^{\rm loc, \,  CR}  \simeq -1.79 \, .
\end{align}

Moreover, the cosines are always large particularly summing only over configurations in this limit. Meaning that for the scales we are considering here all these bispectra are very degenerate in $z\ell m$-space.

We conclude that to attain a precision of $\Delta{\fnlloc} = {\mathcal O}(1)$, it is necessary to take these terms into account. Of course there are many terms in the full expression and a more careful analysis becomes necessary. 
It should be noted that we performed a 2D analysis, fixing the redshift. In principle the redshift evolution can help distinguish the PNG bispectrum from those generated by relativistic effects. We defer a comprehensive analysis including all the dominant contributions from relativistic effects in several redshift bins to a future work. In this work we only show that results do not change qualitatively when considering other redshifts, see e.g.~top and bottom panels of Fig.~\ref{fig:fnllocall}. We remark that at high redshift the non-Gaussian contamination induced by relativistic effects is smaller. Interestingly the contribution of the ``Newtonian~$\times$~Lensing" terms is  relatively larger for higher redshifts.
The two groups of three-derivative terms are characterised by a different behaviour with redshift. Indeed the corrections induced by the group involving peculiar velocities are less relevant at high redshift, as expected, given that some of them inversely proportional to the comoving distance of the sources.

%%%%%%%%%%%%%
\subsection{Equilateral and orthogonal non-Gaussianity}\label{equilng}
\begin{figure}[ht!]
\begin{center}
\includegraphics[width=0.45\textwidth]{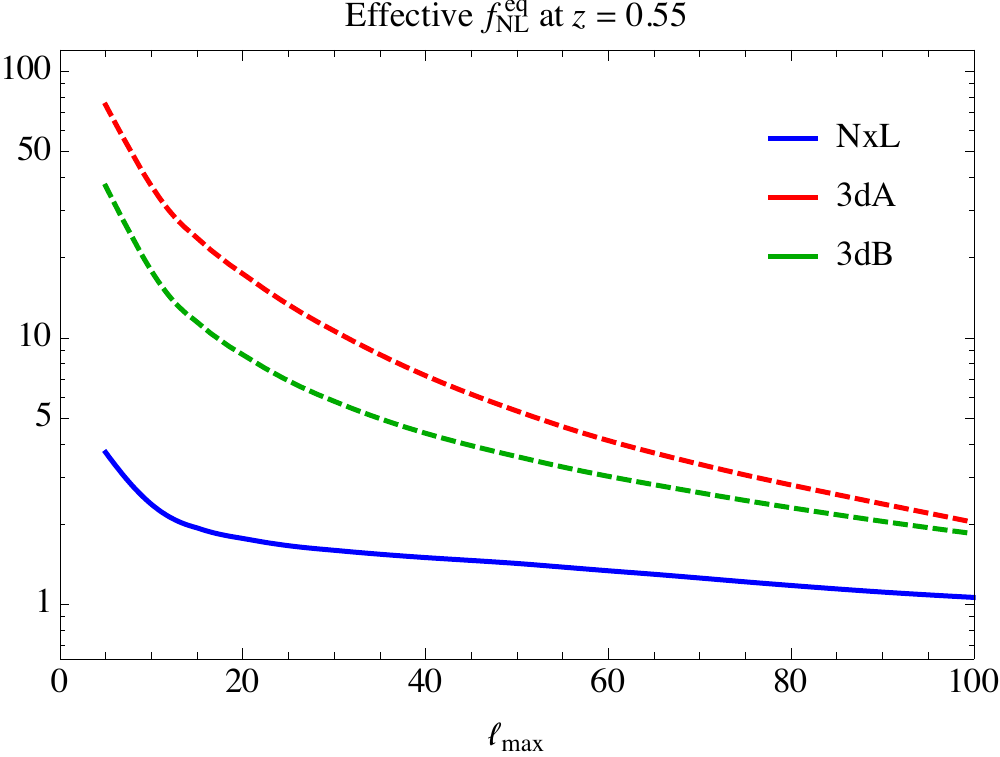} 
\includegraphics[width=0.45\textwidth]{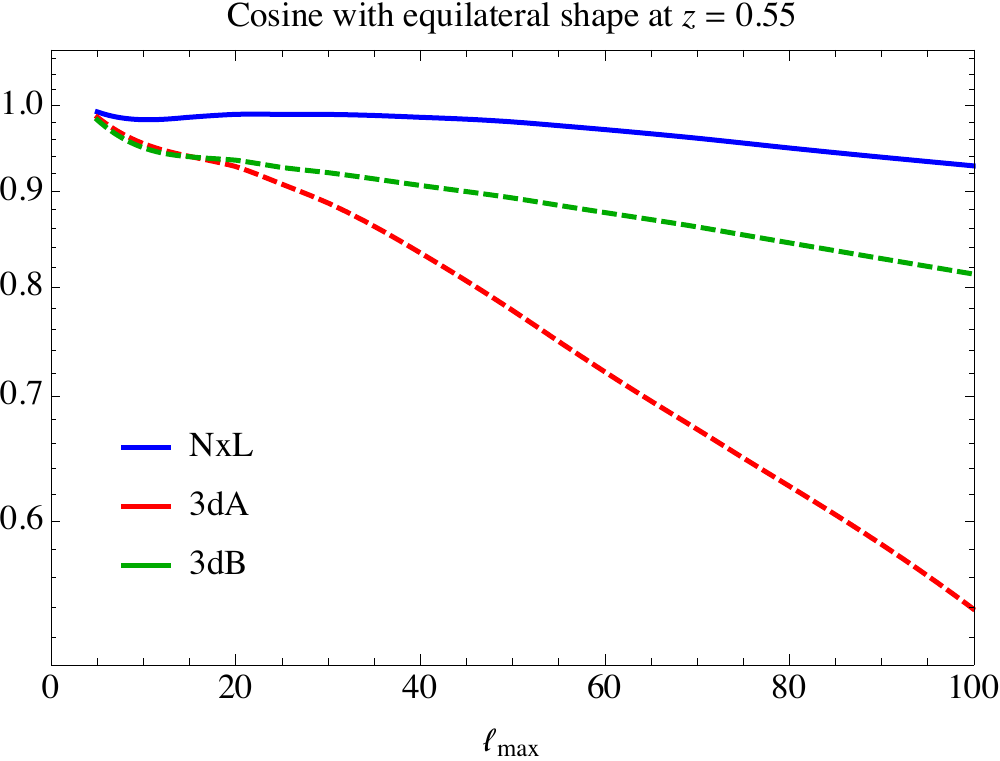}
\caption{Effective equilateral non-Gaussianity $\fnleq$ (left) and cosine with equilateral shape (right) as a function of the maximum $\ell$ included in the configurations at equal redshift $z=0.55$. Dashed lines indicate negative values.
}
\label{fig:fnleqall}
\end{center}
\end{figure}

For the ``Newtonian $\times$ Lensing"  and the ``three-derivative" terms, we repeat the same calculation as above for the equilateral shape. The results are presented in Figure \ref{fig:fnleqall}. We find that for $\ell_{\rm max} = 100$ they lead to a contamination of
\begin{align}
&\fnl^{\rm eq, \, N\times L}   \simeq 1.1, \\
&\fnl^{\rm eq, \, 3dA}  \simeq -2.0, \\
&\fnl^{\rm eq, \, 3dB}  \simeq -1.8\,.
\end{align}
For lower $\ell_{\rm max}$ the contamination is bigger as there is not enough information to lift the degeneracy between the terms and the template. We interpret this as the fact that by including less configurations the peculiarity of the shape is weaker and the overlap with the terms above is therefore greater.

Regarding the cosine between these contributions and the equilateral shape, it can be surprising that the cosine is larger for the ``Newtonian $\times$ Lensing" terms which are the ones producing the smallest contamination to $\fnleq$. This is due to the fact that the amplitude of the ``three-derivative" terms is larger than the one of the Newtonian $\times$ Lensing terms.

%%%%%%%%%%%%%%%%%%%%%%%%%%%%%%%%%%

\begin{figure}[ht!]
\begin{center}
\includegraphics[width=0.45\textwidth]{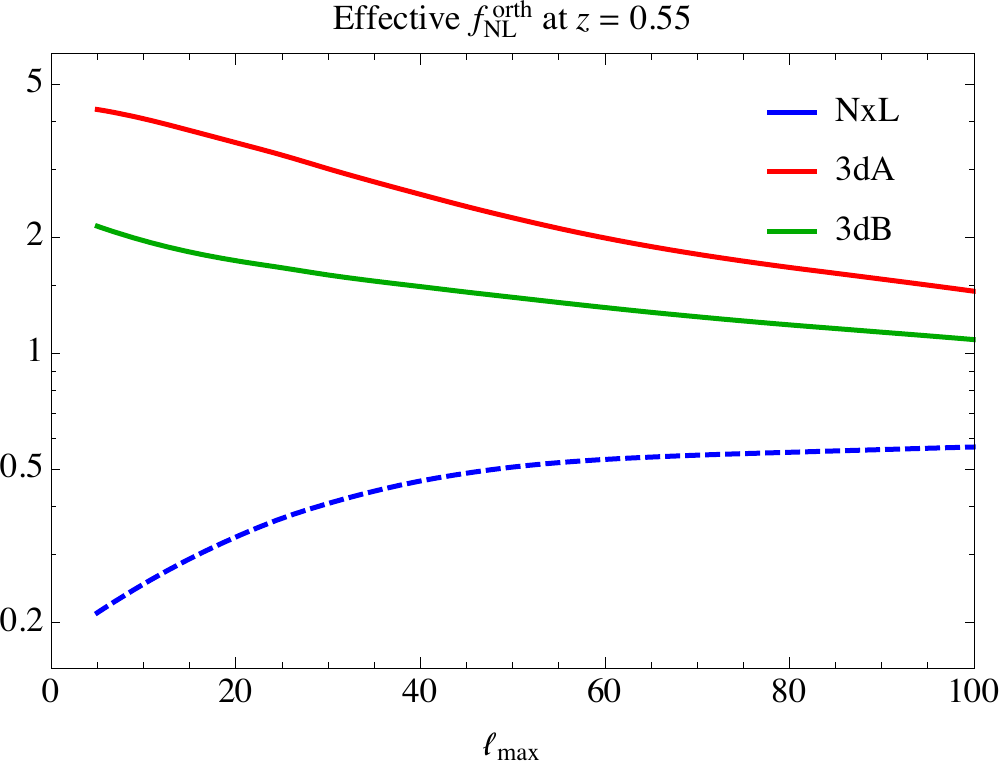} 
\includegraphics[width=0.45\textwidth]{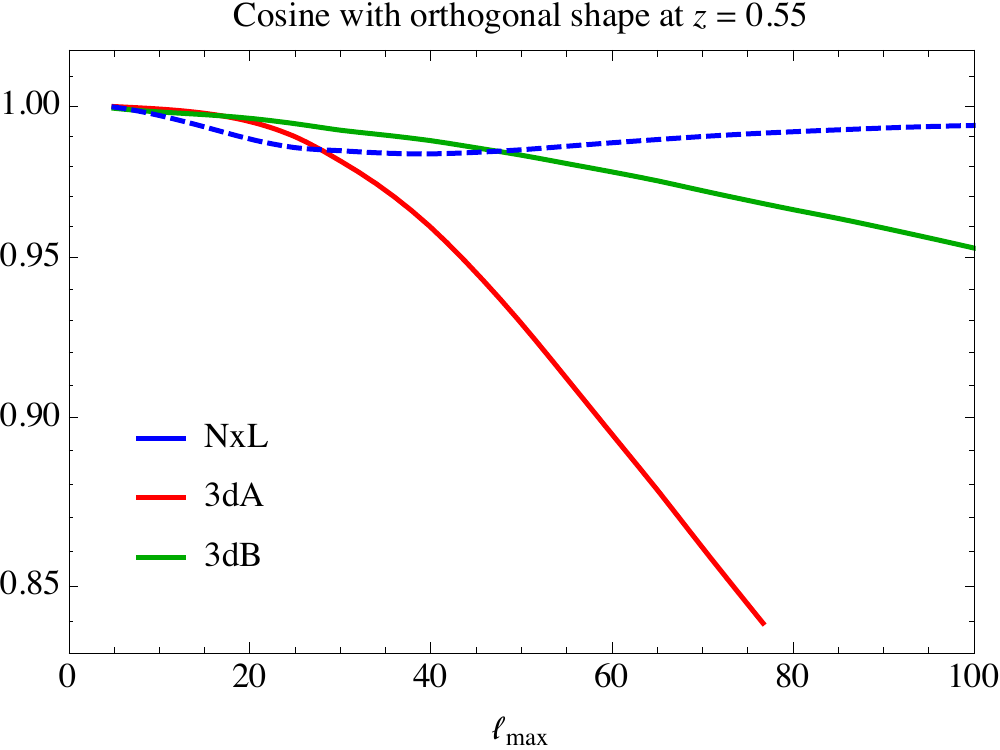}
\caption{Effective orthogonal non-Gaussianity $\fnlorth$ (left) and cosine with orthogonal shape (right) as a function of the maximum $\ell$ included in the configurations at equal redshift $z=0.55$. Dashed lines indicate negative values.
}
\label{fig:fnlorthall}
\end{center}
\end{figure}

We also compute the effective orthogonal non-Gaussianity induced by the subset of projection effects considered above, i.e.~``N$\times$L'' and ``three-derivative'' terms. The results are shown in Fig.~\ref{fig:fnlorthall}. At $\ell_{\rm max} = 100$ they lead to a contamination of
\begin{align}
&\fnl^{\rm orth, \, N\times L}   \simeq -0.57, \\
&\fnl^{\rm orth, \, 3dA}  \simeq 1.5, \\
&\fnl^{\rm orth, \, 3dB}  \simeq 1.1\,.
\end{align}

We find that the contamination on the primordial signal of a type given by the equilateral and orthogonal templates is below the sensitivity of upcoming surveys.

%%%%%%%%%%%%%%%%%%%%%%%%%%%%%%%%%%%
\section{Conclusions}\label{secconclusion}

In addition to the contribution from primordial non-Gaussianity, there are several other sources of non-Gaussianity that give rise to non-zero bispectrum and hence should be modelled correctly in order to extract the primordial signal. On small scales, modelling of the non-linear gravitational evolution of matter, non-linearity in biasing relation and the non-linear redshift space distortions is necessary, while on large scales or when correlating different redshift bins, a fully general relativistic calculation of the bispectrum is also needed. Our focus in this paper has been on the general relativistic corrections to the bispectrum due to projection effects.

We have computed the effective non-Gaussianity imprinted from relativistic corrections to the bispectrum of galaxies from several terms appearing on the explicit computation of the relativistic projection effects. This should be understood as a systematic error that is induced in the estimated primordial signal by such terms if they are ignored. By estimating the order of magnitude of these effects, we addressed the question of whether they can be safely ignored in future analyses. We developed a numerical scheme to calculate the non-Gaussian shape templates for the observed galaxy bispectrum in angular and redshift coordinates. We estimated the systematic shift of the amplitude of the primordial non-Gaussianity $f_{\rm NL}$ parameter for each shape by projecting the theoretical bispectrum onto the corresponding shape and summing over triangular configurations. 

At $z=0.55$ the terms from second-order perturbation theory called Newtonian $\times$ Lensing produce  
a contamination of magnitude 
$\fnlloc \simeq 0.19$ for the local shape, $\fnleq \simeq 1.1$ for equilateral shape, and 
$\fnl^{\rm orth.}   \simeq -0.57$ for the orthogonal shape. The sum of the three-derivative terms we considered  produce a value of $\fnlloc \simeq -0.85$, $\fnleq \simeq -3.8$ and $\fnl^{\rm orth.}   \simeq 2.6$. These are relevant contaminations as they are only a small subset of all the second-order terms. Since they give contributions of the order $\fnl \sim 1$ on the estimated local-type non-Gaussianity, which is the goal to be attained with LSS data, these effects cannot be ignored in the future. Interestingly, we found that in $z\ell m$-space, at a single redshift and with the scales which we have considered, the different shapes have a very significant overlap and it will be important to include both, higher values of $\ell$ and different redshifts in order to distinguish them.
    
 The $\fnl$ found in~\cite{Umeh:2016nuh} is larger than our result. But even though they do consider a different set of terms than we do, we believe that this is mainly a consequence of their normalisation with respect to  the \emph{Newtonian} power spectrum and therefore not a true difference.
    
A full answer to the question addressed in this work requires its extension in several directions. In an estimation of non-Gaussianity from real data, one would need of course to include \textit{all} terms from second order galaxy number counts as there could be combinations  leading to higher or lower level of contamination. It would be interesting to study how to distinguish these contributions from the primordial signal. One possibility is to correlate different redshift bins, but in this case there will be several projection effects that can become larger than the ones we have considered here, such as other lensing contributions. Another interesting question is to give a similar estimate when the data is transformed to Fourier space including integrated terms which can potentially give a large contribution. Finally, since mildly non-linear scales can potentially give an important contribution to the signal, it will be important to compute the galaxy bispectrum beyond tree level (i.e.~beyond second order in perturbation theory). The bispectrum probes the coupling of different scales, and there can be a signal of interesting physics when large and small scales are combined in the squeezed limit, which requires a calculation which is both non-linear and relativistic. Higher-order relativistic calculations quickly become too complicated to be practical. One possibility to solve this issue is to use relativistic numerical simulations \cite{Adamek:2015eda}. Another possibility is that, since relativistic corrections remain small, one can in principle do a mixed perturbation theory where relativistic corrections are kept at low orders, but one goes to higher orders in the quantities that become large at those scales, such as the density perturbation.
\vskip 0.2cm

\acknowledgments{We would like to thank R.~Maartens and O.~Umeh for useful conversations. E.~D. is supported by the ERC Starting Grant cosmoIGM and by INFN/PD51 INDARK grant.
G.M.  wishes to thank CNPq for financial support.
A.M. is supported by the Tomalla foundation for Gravity Research for part of this research.
J.N. is supported by Proyecto VRIEA-PUCV 039.362/2016. A.R. is supported by
the Swiss National Science Foundation (SNSF), project {\it Investigating the Nature of Dark
Matter, project} number: 200020-159223. Also R.D. and H.P. acknowledge support from the Swiss National Science Foundation.
We thank the Galileo Galilei Institute for Theoretical Physics for the hospitality and the INFN for partial support during an intermediate stage of this project.
}

\appendix
%%%%%%%%%%%%%%%%%%%%%%%%%%%%%%%%%%%%%
\section{Three-derivative bispectrum}
\label{sec:3diff_appendix}
%%%%%%%%%%%%%%%%%%%%%%%%%%%%%%%%%%%%%
In this appendix, by following the notation of Ref.~\cite{DiDio:2015bua}, we compute the redshift dependent bispectra for the terms with three spatial derivatives defined through Eqs.~(\ref{3diff}) and (\ref{e:SIS-3der-B}). 
%%%%%%
\subsection{Term $\delta \partial_r v $}
\label{sec:B1}
We compute the contribution of the following term.
\be
\langle \left(  \delta \partial_r v \right) \left( \bn_1 , z_1 \right) \delta \left( \bn_2 ,z_2 \right)  \delta \left( \bn_3 ,z_3 \right) \rangle_c + \text{perms.}, 
\ee
finding
\bea
\langle ... \rangle_c &=&
\frac{1}{ \left( 2\pi\right)^{12}} \int d^3 k d^3k_1 d^3k_2 d^3k_3 \,T_\delta \left( k, \eta \right) T_v \left( k_1 , \eta \right) T_\delta \left( k_2, \eta_2 \right) T_\delta \left( k_3, \eta_3 \right)  
\nonumber \\
&&
\qquad \times
\langle \zeta \left( \bk \right) \zeta \left( \bk_1 \right) \zeta \left( \bk_2 \right) \zeta \left( \bk_3 \right) \rangle_c \left( \partial_{k_1 r_1} e^{i \bk_1 \cdot \bn_1 r_1 } \right)   e^{i \left( \bk \cdot\bn_1 r_1 + \bk_2 \cdot \bn_2 r_2+ \bk_3 \cdot \bn_3 r_3\right)} 
\nonumber \\
&=&
\frac{1}{ \left( 2\pi\right)^{6}} \int d^3k_2 d^3k_3 \, T_\delta \left( k_2, \eta \right) T_v \left( k_3 , \eta \right) T_\delta \left( k_2, \eta_2 \right) T_\delta \left( k_3, \eta_3 \right)  P_\zeta \left( k_2 \right) P_\zeta\left( k_3 \right)
\nonumber \\
&&
\qquad \times
\left( \partial_{ k_3 r_1  } e^{-i \bk_3\cdot \bn_1 r_1} \right)   e^{i \left( -\bk_2\cdot  \bn_1 r_1+ \bk_2 \cdot \bn_2  r_2+ \bk_3 \cdot \bn_3 r_3\right)} 
\nonumber \\
&+& 
\frac{1}{ \left( 2\pi\right)^{6}} \int d^3k_2 d^3k_3 \, T_\delta \left( k_3, \eta \right) T_v \left( k_2 , \eta \right) T_\delta \left( k_2, \eta_2 \right) T_\delta \left( k_3, \eta_3 \right)  P_\zeta\left( k_2 \right) P_\zeta\left( k_3 \right)
\nonumber \\
&&
\qquad \times
\left( \partial_{ k_2 r_1  } e^{-i \bk_2 \cdot \bn_1 r_1 } \right)   e^{i \left( -\bk_3 \cdot \bn_1 r_1 + \bk_2 \cdot \bn_2 r_2+ \bk_3 \cdot \bn_3 r_3 \right)} \, .
\eea
Then, by expanding the exponentials in spherical harmonics and spherical Bessel functions, we obtain
\bea
B^{\delta v} \left( \bn_1 , \bn_2 , \bn_3 , z_1, z_2 , z_3 \right) &=& \frac{4}{\pi^2} \sum_{\substack{\ell, \ell'\\ m, m'}} Y_{\ell m} \left( \bn_1 \right) Y_{\ell' m'} \left( \bn_1 \right) Y^*_{\ell m } \left( \bn_2 \right) Y^*_{\ell' m'}\left( \bn_3 \right) Z^{\delta v}_{\ell \ell'} \left( z_1 , z_2 ,z_3 \right)
\nonumber
\\
&&
+ \text{perms.} \,,
\eea
with
\bea
Z^{\delta v}_{\ell \ell'} \left( z_1 , z_2 ,z_3 \right) &=&
\int dk_2 dk_3  k_2^2 k_3^2 \,T_\delta \left( k_2, \eta \right) T_v \left( k_3 , \eta \right) T_\delta \left( k_2, \eta_2 \right) T_\delta \left( k_3, \eta_3 \right)  
\nonumber \\
&&
\times P_\zeta\left( k_2 \right) P_\zeta\left( k_3 \right) j_\ell \left( k_2 r_1 \right)  j'_{\ell'} \left( k_3 r_1\right)  j_\ell \left( k_2 r_2 \right)  j_{\ell'} \left( k_3 r_3\right) 
\nonumber \\
&+&
\int dk_2 dk_3  k_2^2 k_3^2 \,T_\delta \left( k_3, \eta \right) T_v \left( k_2 , \eta \right) T_\delta \left( k_2, \eta_2 \right) T_\delta \left( k_3, \eta_3 \right)  
\nonumber \\
&&
\times P_\zeta\left( k_2 \right) P_\zeta\left( k_3 \right) j'_\ell \left( k_2 r_1 \right)  j_{\ell'} \left( k_3 r_1\right)  j_\ell \left( k_2 r_2 \right)  j_{\ell'} \left( k_3 r_3\right) \,.
\eea
The reduced bispectrum is then given by
\be
b^{\delta v}_{\ell_1 \ell_2 \ell_3} \left( z_1 , z_2 , z_3 \right) = \frac{4}{\pi^2} Z^{v\delta}_{\ell_2 \ell_3}  \left( z_1 , z_2 , z_3 \right) + \text{perms.}
\ee
and in terms of angular power spectra by
\be
b^{\delta v}_{\ell_1 \ell_2 \ell_3} \left( z_1 , z_2 , z_3 \right) = C_{\ell_2}^{\delta \delta} \left( z_1, z_2 \right)  C_{\ell_3}^{v \delta} \left( z_1, z_3 \right) 
+C_{\ell_3}^{\delta \delta} \left( z_1, z_3 \right)  C_{\ell_2}^{v \delta} \left( z_1, z_2 \right) 
+ \text{perms.}\,.
\ee

%%%%%%
\subsection{Term $\HH^{-1} \partial_r v \partial^2_r v $}
\label{sec:B2}
Similarly, we compute 
\be
\langle \left(\HH^{-1} \partial_r v \partial^2_r v  \right) \left( \bn_1 , z_1 \right) \delta \left( \bn_2 ,z_2 \right)  \delta \left( \bn_3 ,z_3 \right) \rangle_c + \text{perms.}, 
\ee
and we find
\bea
\langle ... \rangle_c &=&
\frac{1}{ \HH \left( z_1 \right) \left( 2\pi\right)^{12}} \int d^3 k d^3k_1 d^3k_2 d^3k_3  k_1  T_v \left( k, \eta \right) T_v \left( k_1 , \eta \right) T_\delta \left( k_2, \eta_2 \right) T_\delta \left( k_3, \eta_3 \right)  
\nonumber \\
&&
\qquad \times
\langle \zeta \left( \bk \right) \zeta \left( \bk_1 \right) \zeta \left( \bk_2 \right) \zeta \left( \bk_3 \right) \rangle_c 
\left( \partial_{k r_1} e^{i  \bk \cdot \bn_1 r_1} \right)
\left( \partial_{ k_1 r_1  }^2 e^{i \bk_1 \cdot \bn_1 r_1} \right)   e^{ i \left( \bk_2 \cdot \bn_2 r_2 + \bk_3 \cdot \bn_3 r_3 \right)} 
\nonumber \\
&=&
\frac{1}{\HH \left( z_1 \right) \left( 2\pi\right)^{6}} \int d^3k_2 d^3k_3 k_2  T_v \left( k_2, \eta \right) T_v \left( k_3 , \eta \right) T_\delta \left( k_2, \eta_2 \right) T_\delta \left( k_3, \eta_3 \right)  P_\zeta\left( k_2 \right) 
P_\zeta\left( k_3 \right)
\nonumber \\
&&
\qquad \times
\left( \partial_{ k_2 r_1  }^2 e^{-i \bk_2 \cdot\bn_1 r_1 } \right)  \left( \partial_{ k_3 r_1  } e^{-i \bk_3 \cdot \bn_1 r_1 } \right)   e^{i \left(  \bk_2 \cdot \bn_2 r_2 + \bk_3 \cdot \bn_3 r_3 \right)} 
\nonumber \\
&+& 
\frac{1}{\HH \left( z_1 \right) \left( 2\pi\right)^{6}} \int d^3k_2 d^3k_3 k_3  T_v \left( k_3, \eta \right) T_v \left( k_2 , \eta \right) T_\delta \left( k_2, \eta_2 \right) T_\delta \left( k_3, \eta_3 \right)  P_\zeta\left( k_2 \right) 
P_\zeta\left( k_3 \right)
\nonumber \\
&&
\qquad \times
\left( \partial_{ k_3 r_1 }^2 e^{-i \bk_3\cdot \bn_1 r_1 } \right)   \left( \partial_{ k_2 r_1 } e^{-i \bk_2 \cdot \bn_1 r_1} \right)   e^{i \left(  \bk_2 \cdot \bn_2 r_2+ \bk_3 \cdot \bn_3 r_3\right)} \, .
\eea
Expanding in spherical harmonics we obtain
\be
b^{v'v}_{\ell_1 \ell_2 \ell_3} \left( z_1 , z_2 , z_3 \right) = \frac{4}{\pi^2} Z^{vv'}_{\ell_2 \ell_3}  \left( z_1 , z_2 , z_3 \right) + \text{perms.}\,,
\ee
with
\bea
Z^{v' v}_{\ell \ell'} \left( z_1 , z_2 ,z_3 \right) &=& \frac{1}{\HH \left( z_1 \right)}
\int dk_2 dk_3  k_2^2 k_3^3 \,T_v \left( k_2, \eta \right) T_v \left( k_3 , \eta \right) T_\delta \left( k_2, \eta_2 \right) T_\delta \left( k_3, \eta_3 \right)  
\nonumber \\
&&
\times P_\zeta\left( k_2 \right) P_\zeta\left( k_3 \right) j'_\ell \left( k_2 r_1 \right)  j''_{\ell'} \left( k_3 r_1\right)  j_\ell \left( k_2 r_2 \right)  j_{\ell'} \left( k_3 r_3\right) 
\nonumber \\
&+&
\frac{1}{\HH \left( z_1 \right)}
\int dk_2 dk_3  k_2^3 k_3^2 \,T_v \left( k_3, \eta \right) T_v \left( k_2 , \eta \right) T_\delta \left( k_2, \eta_2 \right) T_\delta \left( k_3, \eta_3 \right)  
\nonumber \\
&&
\times P_\zeta\left( k_2 \right) P_\zeta\left( k_3 \right) j''_\ell \left( k_2 r_1 \right)  j'_{\ell'} \left( k_3 r_1\right)  j_\ell \left( k_2 r_2 \right)  j_{\ell'} \left( k_3 r_3\right) \, .
\eea
It can also be written as a product of power spectra 
\be
b^{v'v}_{\ell_1 \ell_2 \ell_3} \left( z_1 , z_2 , z_3 \right) = C_{\ell_2}^{v \delta} \left( z_1, z_2 \right)  C_{\ell_3}^{v' \delta} \left( z_1, z_3 \right) 
+C_{\ell_3}^{v \delta} \left( z_1, z_3 \right)  C_{\ell_2}^{v' \delta} \left( z_1, z_2 \right) 
+ \text{perms.}\,.
\ee

%%%%%%
\subsection{Term $\HH^{-1} \dot\delta \partial_r v $}
The calculation of
\be
\langle \left( \HH^{-1} \dot \delta \partial_r v \right) \left( \bn_1 , z_1 \right) \delta \left( \bn_2 ,z_2 \right)  \delta \left( \bn_3 ,z_3 \right) \rangle_c + \text{perms.}, 
\ee
can be performed exactly as Sect.~\ref{sec:B1}, simply by replacing $\delta$ with $\HH^{-1} \dot \delta$. Therefore, the final reduced bispectrum can be directly written as
\be
b^{\dot\delta v}_{\ell_1 \ell_2 \ell_3} \left( z_1 , z_2 , z_3 \right) = C_{\ell_2}^{\dot\delta \delta} \left( z_1, z_2 \right)  C_{\ell_3}^{v \delta} \left( z_1, z_3 \right) 
+C_{\ell_3}^{\dot\delta \delta} \left( z_1, z_3 \right)  C_{\ell_2}^{v \delta} \left( z_1, z_2 \right) 
+ \text{perms.} \,.
\ee

%%%%%%
\subsection{Term $\HH^{-2} \Phi_W \partial^3_r v $}
We compute now the bispectrum induced by
\be
\langle \left( \HH^{-2} \Phi_W \partial^3_r v \right) \left( \bn_1 , z_1 \right) \delta \left( \bn_2 ,z_2 \right)  \delta \left( \bn_3 ,z_3 \right) \rangle_c + \text{perms.}\,.
\ee
Therefore, we obtain
\bea
\langle ... \rangle_c &=&
\frac{1}{ 2 \HH^2 \left( z_1 \right) \left( 2\pi\right)^{12}} \int d^3 k d^3k_1 d^3k_2 d^3k_3   k_1^2  \,T_{\Phi + \Psi} \left( k, \eta \right) T_v \left( k_1 , \eta \right) T_\delta \left( k_2, \eta_2 \right) T_\delta \left( k_3, \eta_3 \right)  
\nonumber \\
&&
\qquad \times
\langle \zeta \left( \bk \right) \zeta \left( \bk_1 \right) \zeta \left( \bk_2 \right) \zeta \left( \bk_3 \right) \rangle_c 
\left( \partial_{ k_1 r_1  }^3 e^{i \bk_1 \cdot \bn_1 r_1} \right)   e^{ i \left( \bk \cdot \bn_1 r_1+ \bk_2 \cdot \bn_2 r_2+ \bk_3 \cdot \bn_3 r_3 \right)} 
\nonumber \\
&=&
\frac{1}{2 \HH^2 \left( z_1 \right) \left( 2\pi\right)^{6}} \int d^3k_2 d^3k_3 k_2^2 \,  T_{v} \left( k_2, \eta \right) T_{\Phi+\Psi} \left( k_3 , \eta \right) T_\delta \left( k_2, \eta_2 \right) T_\delta \left( k_3, \eta_3 \right)  
\nonumber \\
&&
\qquad \times
P_\zeta\left( k_2 \right) P_\zeta\left( k_3 \right)
\left( \partial_{ k_2 r_1  }^3 e^{-i \bk_2 \cdot \bn_1 r_1 } \right)   e^{i \left( - \bk_3 \cdot \bn_1 r_1  + \bk_2 \cdot \bn_2 r_2 + \bk_3 \cdot \bn_3 r_3 \right)} 
\nonumber \\
&+& 
\frac{1}{2\HH^2 \left( z_1 \right) \left( 2\pi\right)^{6}} \int d^3k_2 d^3k_3  k_3^2  \,T_v \left( k_3, \eta \right) T_{\Phi + \Psi} \left( k_2 , \eta \right) T_\delta \left( k_2, \eta_2 \right) T_\delta \left( k_3, \eta_3 \right)  
\nonumber \\
&&
\qquad \times
P_\zeta\left( k_2 \right) P_\zeta\left( k_3 \right)
\left( \partial_{ k_3 r_1 }^3 e^{-i \bk_3 \cdot \bn_1 r_1 } \right)      e^{i \left( - \bk_2 \cdot  \bn_1 r_1 + \bk_2 \cdot \bn_2 r_2 + \bk_3 \cdot \bn_3 r_3\right)} \, .
\eea
By expanding in spherical harmonics we find the following reduced bispectrum
\be
b^{\Phi v'' }_{\ell_1 \ell_2 \ell_3} \left( z_1 , z_2 , z_3 \right) = \frac{4}{\pi^2} Z^{\Phi v''}_{\ell_2 \ell_3}  \left( z_1 , z_2 , z_3 \right) + \text{perms.}\,,
\ee
with
\bea
Z^{\Phi v''}_{\ell \ell'} \left( z_1 , z_2 ,z_3 \right) &=& \frac{1}{2 \HH^2 \left( z_1 \right)}
\int dk_2 dk_3  k_2^4 k_3^2 \,T_v \left( k_2, \eta \right) T_{\Phi + \Psi}\left( k_3 , \eta \right) T_\delta \left( k_2, \eta_2 \right) T_\delta \left( k_3, \eta_3 \right)  
\nonumber \\
&&
\times P_\zeta\left( k_2 \right) P_\zeta\left( k_3 \right) j'''_\ell \left( k_2 r_1 \right)  j_{\ell'} \left( k_3 r_1\right)  j_\ell \left( k_2 r_2 \right)  j_{\ell'} \left( k_3 r_3\right) 
\nonumber \\
&+&
\frac{1}{2 \HH^2 \left( z_1 \right)}
\int dk_2 dk_3  k_2^2 k_3^4\, T_v \left( k_3, \eta \right) T_{\Phi + \Psi} \left( k_2 , \eta \right) T_\delta \left( k_2, \eta_2 \right) T_\delta \left( k_3, \eta_3 \right)  
\nonumber \\
&&
\times P_\zeta\left( k_2 \right) P_\zeta\left( k_3 \right) j_\ell \left( k_2 r_1 \right)  j'''_{\ell'} \left( k_3 r_1\right)  j_\ell \left( k_2 r_2 \right)  j_{\ell'} \left( k_3 r_3\right) \,,
\eea
and as a function of power spectra
\be
b^{\Phi v''}_{\ell_1 \ell_2 \ell_3} \left( z_1 , z_2 , z_3 \right) = C_{\ell_2}^{\Phi \delta} \left( z_1, z_2 \right)  C_{\ell_3}^{v'' \delta} \left( z_1, z_3 \right) 
+C_{\ell_3}^{\Phi \delta} \left( z_1, z_3 \right)  C_{\ell_2}^{v'' \delta} \left( z_1, z_2 \right) 
+ \text{perms.} \,.
\ee

%%%%%%
\subsection{Term $\HH^{-2} \partial_r^2 \Phi_W \partial_r v $}
To compute the bispectrum
\be
\langle \left( \HH^{-2} \partial_r^2 \Phi_W \partial_r v \right) \left( \bn_1 , z_1 \right) \delta \left( \bn_2 ,z_2 \right)  \delta \left( \bn_3 ,z_3 \right) \rangle_c + \text{perms.}, 
\ee
it is enough to repeating the calculation of Sect.~\ref{sec:B2}, by replacing $ \partial_r^2 v$ with $ \partial_r^2 \Phi_W$.
Hence, we get straightforward
\be
b^{\Phi'' v}_{\ell_1 \ell_2 \ell_3} \left( z_1 , z_2 , z_3 \right) = C_{\ell_2}^{v \delta} \left( z_1, z_2 \right)  C_{\ell_3}^{\Phi'' \delta} \left( z_1, z_3 \right) 
+C_{\ell_3}^{v \delta} \left( z_1, z_3 \right)  C_{\ell_2}^{\Phi'' \delta} \left( z_1, z_2 \right) 
+ \text{perms.}\,.
\ee

%%%%%%
\subsection{Term $\HH^{-1} \Phi_W \partial_r \delta $}
We end this section by computing 
\be
\langle \left( \HH^{-1} \Phi_W \partial_r \delta \right) \left( \bn_1 , z_1 \right) \delta \left( \bn_2 ,z_2 \right)  \delta \left( \bn_3 ,z_3 \right) \rangle_c + \text{perms.}.
\ee
By expanding the perturbation in Fourier modes we find
\bea
\langle ... \rangle_c &=&
\frac{1}{ 2 \HH \left( z_1 \right) \left( 2\pi\right)^{12}} \int d^3 k d^3k_1 d^3k_2 d^3k_3   k_1 \, T_{\Phi + \Psi} \left( k, \eta \right) T_\delta \left( k_1 , \eta \right) T_\delta \left( k_2, \eta_2 \right) T_\delta \left( k_3, \eta_3 \right)  
\nonumber \\
&&
\qquad \times
\langle \zeta \left( \bk \right) \zeta \left( \bk_1 \right) \zeta \left( \bk_2 \right) \zeta \left( \bk_3 \right) \rangle_c 
\left( \partial_{ k_1 r_1  } e^{i \bk_1 \cdot \bn_1 r_1 } \right)   e^{ i \left( \bk \cdot \bn_1 r_1+ \bk_2 \cdot \bn_2 r_2 + \bk_3 \cdot \bn_3 r_3 \right)} 
\nonumber \\
&=&
\frac{1}{2 \HH \left( z_1 \right) \left( 2\pi\right)^{6}} \int d^3k_2 d^3k_3 k_2  \, T_{\delta} \left( k_2, \eta \right) T_{\Phi+\Psi} \left( k_3 , \eta \right) T_\delta \left( k_2, \eta_2 \right) T_\delta \left( k_3, \eta_3 \right)  
\nonumber \\
&&
\qquad \times
P_\zeta\left( k_2 \right) P_\zeta\left( k_3 \right)
\left( \partial_{ k_2 r_1  } e^{-i \bk_2 \cdot \bn_1 r_1 } \right)   e^{i \left( - \bk_3 \cdot \bn_1 r_1 + \bk_2 \cdot \bn_2 r_2 + \bk_3 \cdot \bn_3 r_3 \right)} 
\nonumber \\
&+& 
\frac{1}{2\HH \left( z_1 \right) \left( 2\pi\right)^{6}} \int d^3k_2 d^3k_3  k_3  \,T_\delta \left( k_3, \eta \right) T_{\Phi + \Psi} \left( k_2 , \eta \right) T_\delta \left( k_2, \eta_2 \right) T_\delta \left( k_3, \eta_3 \right)  
\nonumber \\
&&
\qquad \times
P_\zeta\left( k_2 \right) P_\zeta\left( k_3 \right)
\left( \partial_{ k_3 r_1 } e^{-i \bk_3 \cdot \bn_1 r_1 } \right)      e^{i \left( - \bk_2 \cdot  \bn_1 r_1 + \bk_2 \cdot \bn_2 r_2 + \bk_3 \cdot \bn_3 r_3 \right)} \, .
\eea
Finally, by expanding in spherical harmonics we obtain the reduced bispectrum
\be
b^{\Phi \delta' }_{\ell_1 \ell_2 \ell_3} \left( z_1 , z_2 , z_3 \right) = \frac{4}{\pi^2} Z^{\Phi \delta'}_{\ell_2 \ell_3}  \left( z_1 , z_2 , z_3 \right) + \text{perms.}\,,
\ee
with
\bea
Z^{\Phi \delta'}_{\ell \ell'} \left( z_1 , z_2 ,z_3 \right) &=& \frac{1}{2 \HH \left( z_1 \right)}
\int dk_2 dk_3  k_2^3 k_3^2 \,T_\delta \left( k_2, \eta \right) T_{\Phi + \Psi}\left( k_3 , \eta \right) T_\delta \left( k_2, \eta_2 \right) T_\delta \left( k_3, \eta_3 \right)  
\nonumber \\
&&
\times P_\zeta\left( k_2 \right) P_\zeta\left( k_3 \right) j'_\ell \left( k_2 r_1 \right)  j_{\ell'} \left( k_3 r_1\right)  j_\ell \left( k_2 r_2 \right)  j_{\ell'} \left( k_3 r_3\right) 
\nonumber \\
&+&
\frac{1}{2 \HH \left( z_1 \right)}
\int dk_2 dk_3  k_2^2 k_3^3\, T_\delta \left( k_3, \eta \right) T_{\Phi + \Psi} \left( k_2 , \eta \right) T_\delta \left( k_2, \eta_2 \right) T_\delta \left( k_3, \eta_3 \right)  
\nonumber \\
&&
\times P_\zeta\left( k_2 \right) P_\zeta\left( k_3 \right) j_\ell \left( k_2 r_1 \right)  j'_{\ell'} \left( k_3 r_1\right)  j_\ell \left( k_2 r_2 \right)  j_{\ell'} \left( k_3 r_3\right) \,,
\eea
and as a function of power spectra
\be
b^{\Phi v''}_{\ell_1 \ell_2 \ell_3} \left( z_1 , z_2 , z_3 \right) = C_{\ell_2}^{\Phi \delta} \left( z_1, z_2 \right)  C_{\ell_3}^{\delta' \delta} \left( z_1, z_3 \right) 
+C_{\ell_3}^{\Phi \delta} \left( z_1, z_3 \right)  C_{\ell_2}^{\delta' \delta} \left( z_1, z_2 \right) 
+ \text{perms.} \,.
\ee

%%%%%%%%%%%%%%%%%%%
\section{Explicit expression for the terms in the bispectrum obtained from consistency relation}\label{explicit_fullconsist}
Using the following notation for the perturbed FLRW metric in Poisson Gauge
\be
ds^2 = a^2(\eta)\left[-(1+2\Psi)d\eta^2 + (1-2\Phi)d{ \bx}^2\right],
\ee
the redshift perturbation induced by the long mode, defined to be minus the redshift perturbations in synchronous gauge, is given by
\be
\Delta z = (1+ z)\bigg[\Psi - \mathcal H v + \partial_r v - \int_{\eta_{\cal O}}^{\eta} \mathrm{d}\eta'\,(\dot \Phi+ \dot \Psi)\bigg]\,.
\label{deltaz}
\ee
The corresponding transfer functions are
\begin{equation}
\Delta_\ell^{\Delta z}(z,k)  = \Delta_\ell^{\Delta z,1}(z,k) + \Delta_\ell^{\Delta z,2}(z,k) +\Delta_\ell^{\Delta z,3}(z,k)  +\Delta_\ell^{\Delta z,4}(z,k), 
\end{equation}  
where
\begin{align*}
\Delta_\ell^{\Delta z, 1} &= (1+z)T_\Psi(\eta,k)j_\ell(kr), \\
\Delta_\ell^{\Delta z, 2} &=  -(1+z) \mathcal{H} T_v(\eta,k)j_\ell(kr), \\
\Delta_\ell^{\Delta z, 3} &= (1+z)  k T_v j_\ell'(kr),\\
\Delta_\ell^{\Delta z, 4} &= -(1+z) \int_{\eta_\mathcal{O}}^{\eta} \d \eta' T_{\dot{\Phi} + \dot{\Psi}} (\eta',k) j_\ell(kr(\eta')).
\end{align*}
In Eq.~\eqref{fullconsist} we apply the chain rule on the redshift partial derivative as
\be \label{partialz}
\partial_z =  \frac{1}{\left( 1+z\right) \HH}\left(  -\partial_\eta + \partial_r \right) \, .
\ee
Since the term
\be
 \frac{1}{\left( 1+z\right) \HH}C_{\ell_1} ^{\Delta z \Delta_g} \left( z_1 , z_2 \right) \partial_{r_2} C_{\ell_3}^{\Delta_g \Delta_g}\left(z_2, z_3\right) 
\ee
includes some contributions of standard perturbation theory, see e.g.~\cite{Bernardeau:2001qr}, we consider only the time derivative in Eq.~\eqref{partialz}.

The volume perturbation at linear order is 
\bea
\frac{\delta V}{V} &= & -2 (\Phi+ \Psi )  + \frac{1}{\HH} \dot\Phi  +\left( \frac{\dot\HH}{{\dot\HH}^2}+\frac{2}{r \mathcal{H}}\right)	\Psi \nonumber \\
& + & \left(-3+\frac{\dot\HH}{\mathcal{H}^2} +\frac{2}{r \mathcal{H}}\right) \left(	\partial_r v - \int_{\eta_{\cal  O}}^\eta \d\eta' (	\dot\Phi + \dot\Psi )  \right) \nonumber  \\
&  - & \int_{\eta_{\cal  O}}^\eta\d\eta'  \left(\frac{2}{r'}\right) (\Phi+ \Psi) \nonumber \\
& + & \int_{\eta_{\cal  O}}^\eta\d\eta'  \left(\frac{r -r'}{r' r}\right) \Delta_\Omega (\Phi+ \Psi) \, .
\eea
Here we have neglected second derivatives acting on the long mode. However, we keep the lensing term, even if it contains a second angular derivative acting on the long mode, since a second angular derivative acting on a constant gradient is non-zero.

The corresponding transfer functions are
\bea
\Delta_\ell^{\delta V }(z,k)  &=& \Delta_\ell^{\delta V,1}(z,k) + \Delta_\ell^{\delta V,2}(z,k) +\Delta_\ell^{\delta V,3}(z,k)  +\Delta_\ell^{\delta V, 4}(z,k) +\Delta_\ell^{\delta V,5}(z,k)  
\nonumber \\
& &
+\Delta_\ell^{\delta V,6}(z,k)+\Delta_\ell^{\delta V,7}(z,k), 
\eea
where
\begin{align*}
\Delta_\ell^{\delta V, 1} &= -2 T_{\Phi + \Psi}(\eta,k)  j_\ell(kr), \\
\Delta_\ell^{\delta V, 2} &= \frac{1}{\mathcal{H}} T_{\dot{\Phi}}(\eta,k)  j_\ell(kr), \\
\Delta_\ell^{\delta V, 3} &=  \left(\frac{\dot{\mathcal{H}}}{\mathcal{H}^2} + \frac{2}{r \mathcal{H}} \right) T_{ \Psi}(\eta,k)  j_\ell(kr), \\
\Delta_\ell^{\delta V 4} &=  \left(-3 +\frac{\dot{\mathcal{H}}}{\mathcal{H}^2} + \frac{2}{r \mathcal{H}} \right) k T_v (\eta,k) j_\ell'(kr), \\
\Delta_\ell^{\delta V, 5} &=  - \left(-3 +\frac{\dot{\mathcal{H}}}{\mathcal{H}^2} + \frac{2}{r \mathcal{H}} \right) \int_{\eta_\mathcal{O}}^{\eta} \d \eta' T_{\dot{\Phi}+\dot{\Psi}} (\eta',k) j_\ell(kr'), \\
\Delta_\ell^{\delta V, 6} &= - \int_{\eta_\mathcal{O}}^{\eta} \d \eta' \frac{2}{r'} T_{\Phi+\Psi} (\eta',k) j_\ell(kr') , \\
\Delta_\ell^{\delta V,7} & = \int_{\eta_{\cal  O}}^\eta\d\eta'  \left(\frac{r -r'}{r' r}\right) \ell(\ell+1) T_{\Phi+ \Psi}(\eta',k) j_\ell(kr')\,.
\end{align*}

The luminosity-distance perturbations at first-order is
\bea
\delta \mathcal{D}_l &=& \left(\frac{1}{r \mathcal{H}}-1 \right) \left(\partial_r v + \Psi - \int_{\eta_{\mathcal{O}}}^\eta \d\eta'\, (\dot\Psi + \dot \Phi) \right) \nonumber \\
&-& \int_{\eta_{\mathcal{O}}}^\eta \frac{ \d \eta'}{r'} \left(\Psi + \Phi \right)  - \Phi +  \int_{\eta_{\cal  O}}^\eta\d\eta'  \left(\frac{r -r'}{2 r' r}\right) \Delta_\Omega (\Phi+ \Psi).
\eea
The corresponding transfer functions are
\bea
\Delta_\ell^{\delta D_l }(z,k)  &=& \Delta_\ell^{\delta D_l,1}(z,k) + \Delta_\ell^{\delta D_l,2}(z,k) +\Delta_\ell^{\delta D_l,3}(z,k)  +\Delta_\ell^{\delta D_l,4}(z,k) +\Delta_\ell^{\delta D_l,5}(z,k) 
\nonumber \\ & &
+\Delta_\ell^{\delta D_l,6}(z,k)\,,
\eea
where
\begin{align*}
\Delta_\ell^{\delta D_l, 1} &=\left(\frac{1}{r \mathcal{H}}-1 \right)  k T_v j_\ell'(kr), \\
\Delta_\ell^{\delta D_l, 2} &= \left(\frac{1}{r \mathcal{H}}-1 \right) T_\Psi (\eta,k) j_\ell(kr), \\
\Delta_\ell^{\delta D_l, 3} &=  \left(1 - \frac{1}{r \mathcal{H}} \right) \int_{\eta_\mathcal{O}}^{\eta} \d \eta' T_{\dot{\Phi} + \dot{\Psi}} (\eta',k) j_\ell(kr(\eta')), \\
\Delta_\ell^{\delta D_l, 4} &= -  \int_{\eta_\mathcal{O}}^{\eta} \frac{\d \eta'}{r'} T_{\Phi +\Psi} (\eta',k) j_\ell(kr(\eta')), \\
\Delta_\ell^{\delta D_l, 5} &= - T_\Phi (\eta,k) j_\ell(kr) , \\
\Delta_\ell^{\delta D_l,6} &=  \int_{\eta_{\cal  O}}^\eta\d\eta'  \left(\frac{r -r'}{2 r' r}\right) \ell(\ell+1) T_{\Phi+ \Psi}(\eta',k) j_\ell(kr') \,.
\end{align*}
The quantity $I$ in Eq.~\eqref{fullconsist} is given by
\be
I = \Phi +\left(\mathcal{H} -\frac{1}{r} \right)v+\int_{\eta_{\cal O}}^{\eta}\frac{\d\eta'}{r'}(\Phi + \Psi) \,,
\ee
where $r=\eta_\mathcal{O} - \eta$ and $r'=\eta_\mathcal{O} - \eta'$. 
The corresponding transfer functions are
\begin{equation}
\Delta_\ell^{I}(z,k)  = \Delta_\ell^{I,1}(z,k) + \Delta_\ell^{I,2}(z,k) +\Delta_\ell^{I,3}(z,k), 
\end{equation}  
where
\begin{align*}
\Delta_\ell^{I, 1} &= T_\Phi(\eta,k)j_\ell(kr), \\
\Delta_\ell^{I, 2} &= \left( \mathcal{H} - \frac{1}{r} \right)  T_v j_\ell(kr),\\
\Delta_\ell^{I, 3} &=  \int_{\eta_\mathcal{O}}^{\eta}\frac{ \d \eta'}{r'} T_{\Phi+\Psi} (\eta',k) j_\ell(kr(\eta')).
\end{align*}

%%%%%%%%%%%%%%%%%%%%%%%%%%%%%%%%%%%%%
\section{Systematic shift on $f_{\rm NL}$}
\label{app:systematic}

In this appendix we briefly review why the projection of a secondary bispectrum on a template is a good estimate of the systematic shift obtained if that secondary effect is neglected. Assume that the full signal  is given by $B_{\rm real} = B_{\rm N} + B_{\rm GR} + \bar{f}_{\rm NL} B_{\rm shape}$ where $B_{\rm N}$ is the bispectrum from Newtonian perturbation theory and $B_{\rm GR}$ is the relativistic contributions. Now asume that one wants to extract $f_{\rm NL}$ by comparing with a model which is given by $ B_{\rm model} = B_{\rm N} + f_{\rm NL} B_{\rm shape}$. The log likelihood would be approximately
\begin{align}
\log \mathcal{L} &\supset -\frac{1}{2}\sum (B_{\rm real} - B_{\rm model}) C^{-1} (B_{\rm real} - B_{\rm model})\nn \\
 &= -\frac{1}{2}\sum (B_{\rm GR} + (\bar{f}_{\rm NL} - f_{\rm NL}) B_{\rm shape}) C^{-1} (B_{\rm GR} + (\bar{f}_{\rm NL} - f_{\rm NL}) B_{\rm shape})\,,
\end{align}
where $C$ is the covariance matrix of the full bispectrum. We want to find the extrema of this function in order to get the ``measured'' value of $f_{\rm NL}$ (the one you would think you measure if you ignore relativistic effects). We thus obtain
\be
f_{\rm NL} - \bar{f}_{\rm NL}  =\left( \sum B_{\rm shape} C^{-1} B_{\rm GR}\right)/\left(\sum B_{\rm shape} C^{-1} B_{\rm shape}\right)\,.
\ee

\bibliographystyle{JHEP}
\bibliography{biblio_fnl}

\providecommand{\href}[2]{#2}\begingroup\raggedright\begin{thebibliography}{10}

\bibitem{Ade:2015ava}
{\bf Planck} Collaboration, P.~A.~R. Ade et~al., {\it {Planck 2015 results.
  XVII. Constraints on primordial non-Gaussianity}},
  \href{http://arxiv.org/abs/1502.01592}{{\tt arXiv:1502.01592}}.

\bibitem{Bartolo:2004if}
N.~Bartolo, E.~Komatsu, S.~Matarrese, and A.~Riotto, {\it {Non-Gaussianity from
  inflation: Theory and observations}},  {\em Phys. Rept.} {\bf 402} (2004)
  103--266, [\href{http://arxiv.org/abs/astro-ph/0406398}{{\tt
  astro-ph/0406398}}].

\bibitem{Dore:2014cca}
O.~Dor{\'e} et~al., {\it {Cosmology with the SPHEREX All-Sky Spectral Survey}},
   \href{http://arxiv.org/abs/1412.4872}{{\tt arXiv:1412.4872}}.

\bibitem{Tellarini:2016sgp}
M.~Tellarini, A.~J. Ross, G.~Tasinato, and D.~Wands, {\it {Galaxy bispectrum,
  primordial non-Gaussianity and redshift space distortions}},  {\em JCAP} {\bf
  1606} (2016) 014, [\href{http://arxiv.org/abs/1603.06814}{{\tt
  arXiv:1603.06814}}].

\bibitem{Baldauf:2016sjb}
T.~Baldauf, M.~Mirbabayi, M.~Simonovi{\'c}, and M.~Zaldarriaga, {\it {LSS
  constraints with controlled theoretical uncertainties}},
  \href{http://arxiv.org/abs/1602.00674}{{\tt arXiv:1602.00674}}.

\bibitem{Dalal:2007cu}
N.~Dalal, O.~Dore, D.~Huterer, and A.~Shirokov, {\it {The imprints of
  primordial non-gaussianities on large-scale structure: scale dependent bias
  and abundance of virialized objects}},  {\em Phys. Rev.} {\bf D77} (2008)
  123514, [\href{http://arxiv.org/abs/0710.4560}{{\tt arXiv:0710.4560}}].

\bibitem{Matarrese:2008nc}
S.~Matarrese and L.~Verde, {\it {The effect of primordial non-Gaussianity on
  halo bias}},  {\em Astrophys. J.} {\bf 677} (2008) L77--L80,
  [\href{http://arxiv.org/abs/0801.4826}{{\tt arXiv:0801.4826}}].

\bibitem{Afshordi:2008ru}
N.~Afshordi and A.~J. Tolley, {\it {Primordial non-gaussianity, statistics of
  collapsed objects, and the Integrated Sachs-Wolfe effect}},  {\em Phys. Rev.}
  {\bf D78} (2008) 123507, [\href{http://arxiv.org/abs/0806.1046}{{\tt
  arXiv:0806.1046}}].

\bibitem{Slosar:2008hx}
A.~Slosar, C.~Hirata, U.~Seljak, S.~Ho, and N.~Padmanabhan, {\it {Constraints
  on local primordial non-Gaussianity from large scale structure}},  {\em JCAP}
  {\bf 0808} (2008) 031, [\href{http://arxiv.org/abs/0805.3580}{{\tt
  arXiv:0805.3580}}].

\bibitem{Scoccimarro:2003wn}
R.~Scoccimarro, E.~Sefusatti, and M.~Zaldarriaga, {\it {Probing primordial
  non-Gaussianity with large - scale structure}},  {\em Phys. Rev.} {\bf D69}
  (2004) 103513, [\href{http://arxiv.org/abs/astro-ph/0312286}{{\tt
  astro-ph/0312286}}].

\bibitem{Sefusatti:2007ih}
E.~Sefusatti and E.~Komatsu, {\it {The bispectrum of galaxies from
  high-redshift galaxy surveys: Primordial non-Gaussianity and non-linear
  galaxy bias}},  {\em Phys. Rev.} {\bf D76} (2007) 083004,
  [\href{http://arxiv.org/abs/0705.0343}{{\tt arXiv:0705.0343}}].

\bibitem{Jeong:2009vd}
D.~Jeong and E.~Komatsu, {\it {Primordial non-Gaussianity, scale-dependent
  bias, and the bispectrum of galaxies}},  {\em Astrophys. J.} {\bf 703} (2009)
  1230--1248, [\href{http://arxiv.org/abs/0904.0497}{{\tt arXiv:0904.0497}}].

\bibitem{Lewis:2012tc}
A.~Lewis, {\it {The full squeezed CMB bispectrum from inflation}},  {\em JCAP}
  {\bf 1206} (2012) 023, [\href{http://arxiv.org/abs/1204.5018}{{\tt
  arXiv:1204.5018}}].

\bibitem{Kim:2013nea}
J.~Kim, A.~Rotti, and E.~Komatsu, {\it {Removing the ISW-lensing bias from the
  local-form primordial non-Gaussianity estimation}},  {\em JCAP} {\bf 1304}
  (2013) 021, [\href{http://arxiv.org/abs/1302.5799}{{\tt arXiv:1302.5799}}].

\bibitem{Yoo:2010ni}
J.~Yoo, {\it {General Relativistic Description of the Observed Galaxy Power
  Spectrum: Do We Understand What We Measure?}},  {\em Phys.Rev.} {\bf D82}
  (2010) 083508, [\href{http://arxiv.org/abs/1009.3021}{{\tt
  arXiv:1009.3021}}].

\bibitem{Bonvin:2011bg}
C.~Bonvin and R.~Durrer, {\it {What galaxy surveys really measure}},  {\em
  Phys.Rev.} {\bf D84} (2011) 063505,
  [\href{http://arxiv.org/abs/1105.5280}{{\tt arXiv:1105.5280}}].

\bibitem{Challinor:2011bk}
A.~Challinor and A.~Lewis, {\it {The linear power spectrum of observed source
  number counts}},  {\em Phys.Rev.} {\bf D84} (2011) 043516,
  [\href{http://arxiv.org/abs/1105.5292}{{\tt arXiv:1105.5292}}].

\bibitem{Jeong:2011as}
D.~Jeong, F.~Schmidt, and C.~M. Hirata, {\it {Large-scale clustering of
  galaxies in general relativity}},  {\em Phys. Rev.} {\bf D85} (2012) 023504,
  [\href{http://arxiv.org/abs/1107.5427}{{\tt arXiv:1107.5427}}].

\bibitem{Bartolo:2005xa}
N.~Bartolo, S.~Matarrese, and A.~Riotto, {\it {Signatures of primordial
  non-Gaussianity in the large-scale structure of the Universe}},  {\em JCAP}
  {\bf 0510} (2005) 010, [\href{http://arxiv.org/abs/astro-ph/0501614}{{\tt
  astro-ph/0501614}}].

\bibitem{Bartolo:2010rw}
N.~Bartolo, S.~Matarrese, O.~Pantano, and A.~Riotto, {\it {Second-order matter
  perturbations in a LambdaCDM cosmology and non-Gaussianity}},  {\em Class.
  Quant. Grav.} {\bf 27} (2010) 124009,
  [\href{http://arxiv.org/abs/1002.3759}{{\tt arXiv:1002.3759}}].

\bibitem{Villa:2014foa}
E.~Villa, L.~Verde, and S.~Matarrese, {\it {General relativistic corrections
  and non-Gaussianity in large scale structure}},  {\em Class. Quant. Grav.}
  {\bf 31} (2014) 234005, [\href{http://arxiv.org/abs/1409.4738}{{\tt
  arXiv:1409.4738}}].

\bibitem{Verde:1999ij}
L.~Verde, L.-M. Wang, A.~Heavens, and M.~Kamionkowski, {\it {Large scale
  structure, the cosmic microwave background, and primordial non-gaussianity}},
   {\em Mon. Not. Roy. Astron. Soc.} {\bf 313} (2000) L141--L147,
  [\href{http://arxiv.org/abs/astro-ph/9906301}{{\tt astro-ph/9906301}}].

\bibitem{Adamek:2015eda}
J.~Adamek, D.~Daverio, R.~Durrer, and M.~Kunz, {\it {General relativity and
  cosmic structure formation}},  {\em Nature Phys.} {\bf 12} (2016) 346--349,
  [\href{http://arxiv.org/abs/1509.01699}{{\tt arXiv:1509.01699}}].

\bibitem{Bertacca:2014dra}
D.~Bertacca, R.~Maartens, and C.~Clarkson, {\it {Observed galaxy number counts
  on the lightcone up to second order: I. Main result}},  {\em JCAP} {\bf 1409}
  (2014) 037, [\href{http://arxiv.org/abs/1405.4403}{{\tt arXiv:1405.4403}}].

\bibitem{Yoo:2014sfa}
J.~Yoo and M.~Zaldarriaga, {\it {Beyond the Linear-Order Relativistic Effect in
  Galaxy Clustering: Second-Order Gauge-Invariant Formalism}},  {\em Phys.Rev.}
  {\bf D90} (2014) 023513, [\href{http://arxiv.org/abs/1406.4140}{{\tt
  arXiv:1406.4140}}].

\bibitem{DiDio:2014lka}
E.~Di~Dio, R.~Durrer, G.~Marozzi, and F.~Montanari, {\it {Galaxy number counts
  to second order and their bispectrum}},  {\em JCAP} {\bf 1412} (2014) 017,
  [\href{http://arxiv.org/abs/1407.0376}{{\tt arXiv:1407.0376}}]. [Erratum:
  \emph{ JCAP} {\bf 1506}, E01 (2015)].

\bibitem{Kehagias:2015tda}
A.~Kehagias, A.~M. Dizgah, J.~Nore{\~n}a, H.~Perrier, and A.~Riotto, {\it {A
  Consistency Relation for the Observed Galaxy Bispectrum and the Local
  non-Gaussianity from Relativistic Corrections}},  {\em JCAP} {\bf 1508}
  (2015) 018, [\href{http://arxiv.org/abs/1503.04467}{{\tt arXiv:1503.04467}}].

\bibitem{Umeh:2016nuh}
O.~Umeh, S.~Jolicoeur, R.~Maartens, and C.~Clarkson, {\it {A general
  relativistic signature in the galaxy bispectrum}},
  \href{http://arxiv.org/abs/1610.03351}{{\tt arXiv:1610.03351}}.

\bibitem{DiDio:2015bua}
E.~Di~Dio, R.~Durrer, G.~Marozzi, and F.~Montanari, {\it {The bispectrum of
  relativistic galaxy number counts}},  {\em JCAP} {\bf 1601} (2016) 016,
  [\href{http://arxiv.org/abs/1510.04202}{{\tt arXiv:1510.04202}}].

\bibitem{Asorey:2012rd}
J.~Asorey, M.~Crocce, E.~Gaztanaga, and A.~Lewis, {\it {Recovering 3D
  clustering information with angular correlations}},  {\em Mon. Not. Roy.
  Astron. Soc.} {\bf 427} (2012) 1891,
  [\href{http://arxiv.org/abs/1207.6487}{{\tt arXiv:1207.6487}}].

\bibitem{DiDio:2013sea}
E.~Di~Dio, F.~Montanari, R.~Durrer, and J.~Lesgourgues, {\it {Cosmological
  Parameter Estimation with Large Scale Structure Observations}},  {\em JCAP}
  {\bf 1401} (2014) 042, [\href{http://arxiv.org/abs/1308.6186}{{\tt
  arXiv:1308.6186}}].

\bibitem{Chen:2010xka}
X.~Chen, {\it {Primordial Non-Gaussianities from Inflation Models}},  {\em Adv.
  Astron.} {\bf 2010} (2010) 638979,
  [\href{http://arxiv.org/abs/1002.1416}{{\tt arXiv:1002.1416}}].

\bibitem{Liguori:2010hx}
M.~Liguori, E.~Sefusatti, J.~R. Fergusson, and E.~P.~S. Shellard, {\it
  {Primordial non-Gaussianity and Bispectrum Measurements in the Cosmic
  Microwave Background and Large-Scale Structure}},  {\em Adv. Astron.} {\bf
  2010} (2010) 980523, [\href{http://arxiv.org/abs/1001.4707}{{\tt
  arXiv:1001.4707}}].

\bibitem{Gangui:1993tt}
A.~Gangui, F.~Lucchin, S.~Matarrese, and S.~Mollerach, {\it {The Three point
  correlation function of the cosmic microwave background in inflationary
  models}},  {\em Astrophys. J.} {\bf 430} (1994) 447--457,
  [\href{http://arxiv.org/abs/astro-ph/9312033}{{\tt astro-ph/9312033}}].

\bibitem{Wang:1999vf}
L.-M. Wang and M.~Kamionkowski, {\it {The Cosmic microwave background
  bispectrum and inflation}},  {\em Phys. Rev.} {\bf D61} (2000) 063504,
  [\href{http://arxiv.org/abs/astro-ph/9907431}{{\tt astro-ph/9907431}}].

\bibitem{Komatsu:2001rj}
E.~Komatsu and D.~N. Spergel, {\it {Acoustic signatures in the primary
  microwave background bispectrum}},  {\em Phys. Rev.} {\bf D63} (2001) 063002,
  [\href{http://arxiv.org/abs/astro-ph/0005036}{{\tt astro-ph/0005036}}].

\bibitem{Babich:2004gb}
D.~Babich, P.~Creminelli, and M.~Zaldarriaga, {\it {The Shape of
  non-Gaussianities}},  {\em JCAP} {\bf 0408} (2004) 009,
  [\href{http://arxiv.org/abs/astro-ph/0405356}{{\tt astro-ph/0405356}}].

\bibitem{Senatore:2009gt}
L.~Senatore, K.~M. Smith, and M.~Zaldarriaga, {\it {Non-Gaussianities in Single
  Field Inflation and their Optimal Limits from the WMAP 5-year Data}},  {\em
  JCAP} {\bf 1001} (2010) 028, [\href{http://arxiv.org/abs/0905.3746}{{\tt
  arXiv:0905.3746}}].

\bibitem{DiDio:2013bqa}
E.~Di~Dio, F.~Montanari, J.~Lesgourgues, and R.~Durrer, {\it {The CLASSgal code
  for Relativistic Cosmological Large Scale Structure}},  {\em JCAP} {\bf 1311}
  (2013) 044, [\href{http://arxiv.org/abs/1307.1459}{{\tt arXiv:1307.1459}}].

\bibitem{Baldauf:2011bh}
T.~Baldauf, U.~Seljak, L.~Senatore, and M.~Zaldarriaga, {\it {Galaxy Bias and
  non-Linear Structure Formation in General Relativity}},  {\em JCAP} {\bf
  1110} (2011) 031, [\href{http://arxiv.org/abs/1106.5507}{{\tt
  arXiv:1106.5507}}].

\bibitem{bessel_int}
A.~D. Jackson and L.~C. Maximon, {\it {Integrals of Products of Bessel
  Functions}},  {\em SIAM Journal on Mathematical Analysis} {\bf 3} (1972)
  446--460, [\href{http://arxiv.org/abs/http://dx.doi.org/10.1137/0503043}{{\tt
  http://dx.doi.org/10.1137/0503043}}].

\bibitem{Blas:2011rf}
D.~Blas, J.~Lesgourgues, and T.~Tram, {\it {The Cosmic Linear Anisotropy
  Solving System (CLASS) II: Approximation schemes}},  {\em JCAP} {\bf 1107}
  (2011) 034, [\href{http://arxiv.org/abs/1104.2933}{{\tt arXiv:1104.2933}}].

\bibitem{Fergusson:2008ra}
J.~R. Fergusson and E.~P.~S. Shellard, {\it {The shape of primordial
  non-Gaussianity and the CMB bispectrum}},  {\em Phys. Rev.} {\bf D80} (2009)
  043510, [\href{http://arxiv.org/abs/0812.3413}{{\tt arXiv:0812.3413}}].

\bibitem{Lacasa:2011ej}
F.~Lacasa, N.~Aghanim, M.~Kunz, and M.~Frommert, {\it {Characterisation of the
  non-Gaussianity of radio and IR point-sources at CMB frequencies}},  {\em
  Mon. Not. Roy. Astron. Soc.} {\bf 421} (2012) 1982,
  [\href{http://arxiv.org/abs/1107.2251}{{\tt arXiv:1107.2251}}].

\bibitem{Limber:1954zz}
D.~N. Limber, {\it {The Analysis of Counts of the Extragalactic Nebulae in
  Terms of a Fluctuating Density Field. II}},  {\em Astrophys. J.} {\bf 119}
  (1954) 655.

\bibitem{LoVerde:2008re}
M.~LoVerde and N.~Afshordi, {\it {Extended Limber Approximation}},  {\em Phys.
  Rev.} {\bf D78} (2008) 123506, [\href{http://arxiv.org/abs/0809.5112}{{\tt
  arXiv:0809.5112}}].

\bibitem{Nielsen:2016ldx}
J.~T. Nielsen and R.~Durrer, {\it {Higher order relativistic galaxy number
  counts: dominating terms}},  \href{http://arxiv.org/abs/1606.02113}{{\tt
  arXiv:1606.02113}}.

\bibitem{Gasperini:2011us}
M.~Gasperini, G.~Marozzi, F.~Nugier, and G.~Veneziano, {\it {Light-cone
  averaging in cosmology: Formalism and applications}},  {\em JCAP} {\bf 1107}
  (2011) 008, [\href{http://arxiv.org/abs/1104.1167}{{\tt arXiv:1104.1167}}].

\bibitem{Peloso:2013zw}
M.~Peloso and M.~Pietroni, {\it {Galilean invariance and the consistency
  relation for the nonlinear squeezed bispectrum of large scale structure}},
  {\em JCAP} {\bf 1305} (2013) 031, [\href{http://arxiv.org/abs/1302.0223}{{\tt
  arXiv:1302.0223}}].

\bibitem{Kehagias:2013yd}
A.~Kehagias and A.~Riotto, {\it {Symmetries and Consistency Relations in the
  Large Scale Structure of the Universe}},  {\em Nucl. Phys.} {\bf B873} (2013)
  514--529, [\href{http://arxiv.org/abs/1302.0130}{{\tt arXiv:1302.0130}}].

\bibitem{Creminelli:2013mca}
P.~Creminelli, J.~Nore{\~n}a, M.~Simonovi{\'c}, and F.~Vernizzi, {\it
  {Single-Field Consistency Relations of Large Scale Structure}},  {\em JCAP}
  {\bf 1312} (2013) 025, [\href{http://arxiv.org/abs/1309.3557}{{\tt
  arXiv:1309.3557}}].

\bibitem{Bernardeau:2001qr}
F.~Bernardeau, S.~Colombi, E.~Gaztanaga, and R.~Scoccimarro, {\it {Large scale
  structure of the universe and cosmological perturbation theory}},  {\em Phys.
  Rept.} {\bf 367} (2002) 1--248,
  [\href{http://arxiv.org/abs/astro-ph/0112551}{{\tt astro-ph/0112551}}].

\end{thebibliography}\endgroup

\end{document}